\documentclass[12pt]{article}
\pdfoutput=1
\usepackage{geometry,enumerate,amsmath,amssymb}
\usepackage{fullpage}
\usepackage{lscape}
\usepackage{amsthm} 
\usepackage{commath}
\usepackage{color}
\usepackage{graphicx}
\usepackage{caption}
\usepackage{subcaption}
\usepackage{fancyhdr}
\usepackage{rotating}
\usepackage{multirow}
\usepackage{pbox}

\def\edit#1{ #1}

\newcommand{\id}{{1}}
\newcommand{\R}{\mathbb{R}}

\newcommand{\C}{\mathbb{C}}

\newcommand{\GL}{\mathrm{GL}}
\newcommand{\SU}{{SU}}
\newcommand{\SO}{{SO}}

\renewcommand{\O}{\mathrm{O}}

\newcommand{\g}{\mathfrak{g}}
\newcommand{\G}{\mathcal{G}}

\newcommand{\M}{\widehat M}

\DeclareMathOperator{\Tr}{Tr}

\newcommand{\totimes}{\mathop{\tilde \otimes}}

\newcommand{\T}{\mathsf{T}}

\newcommand{\BT}{\mathbb{T}}
\newcommand{\BI}{\mathbb{Y}}
\newcommand{\BY}{\mathbb{Y}}
\newcommand{\BD}{\mathbb{D}}
\newcommand{\BG}{\mathbb{G}}
\newcommand{\Q}{{\BD_2}}

\newcommand{\diag}{\text{diag}}

\newcommand{\be}{\begin{equation}}
\newcommand{\ee}{\end{equation}}

\renewcommand{\tilde}{\widetilde}
\renewcommand{\epsilon}{\varepsilon}

\newcommand{\news}{\setcounter{equation}{0}\quad}
\def\ben{\begin{equation}}
\def\een{\end{equation}}
\def\bea{\begin{eqnarray}}
\def\eea{\end{eqnarray}}

\begin{document}
\title{
\begin{flushright}\ \vskip -2cm {\small {\em \edit{DCPT-13/03}}}\end{flushright}
\vskip 2cm ADHM Polytopes
}
\author{James P.~Allen and
Paul Sutcliffe\\[10pt]
{\em \normalsize Department of Mathematical Sciences, Durham University, Durham DH1 3LE, U.K.}\\[10pt] {\normalsize Email:  \quad 
j.p.allen@durham.ac.uk \quad\&\quad\ p.m.sutcliffe@durham.ac.uk} 
}
\date{\edit{February} 2013}
\maketitle
\begin{abstract}
We discuss the construction of ADHM data for Yang-Mills instantons
with the symmetries of the regular polytopes in four dimensions.
We show that the case of the pentatope can be studied using a simple 
modification of the approach previously developed for platonic data. 
For the remaining polytopes, we describe a framework in which the building
blocks of the ADHM data correspond to the edges in the extended Dynkin
diagram that arises via the McKay correspondence. These building blocks
are then assembled into ADHM data through the identification of pairs of
commuting representations of the associated binary polyhedral group.
We illustrate our procedure by the construction of ADHM data associated
with the pentatope, the hyperoctahedron and the 24-cell, with instanton
charges 4, 7 and 23, respectively. Furthermore, we show that within our
framework these are the lowest possible charges with these symmetries.
Plots of topological charge densities are presented that confirm the
polytope structure and the relation to JNR instanton data is clarified.  

\end{abstract}
\newpage
\section{Introduction}\news
There are many interesting connections between instantons, Skyrmions and 
monopoles (for a review see \cite{book}), including the existence of
solutions with platonic symmetries at the same specific values of the
topological charges. 
Instantons were first related to Skyrmions by Atiyah and Manton 
\cite{AM,AM2} who identified instanton holonomies with Skyrme fields, in
a way that has subsequently found a natural interpretation in the context
of holography \cite{SS,Su1}. This connection is particularly useful because
all instantons can be obtained using purely algebraic means via the
Atiyah-Drinfeld-Hitchin-Manin (ADHM) construction \cite{ADHM}.
Combining the existence of known platonic Skyrmions with the 
Atiyah-Manton description motivated the search for platonic instantons.
The first examples of instantons with tetrahedral and cubic symmetries
were presented in \cite{LM} and a detailed understanding of the 
action of platonic symmetries within the ADHM formulation can be found
in \cite{SiSu}, including the explicit derivation of ADHM data for an
instanton with icosahedral symmetry. 
The applicability of this approach has been
demonstrated by the calculation of some fairly complicated ADHM data
associated with symmetric polyhedra, including the truncated
icosahedron \cite{Su2}.   
Platonic ADHM data has recently found another application in 
generating explicit examples of platonic hyperbolic monopoles \cite{MS},
by making use of an observation of Atiyah \cite{At} that identifies
circle invariant instantons as hyperbolic monopoles.
  
The ADHM construction yields the full $8N$-dimensional moduli space
of $SU(2)$ Yang-Mills instantons of charge $N.$ 
For $N\ge 3$ a subset of dimension $5N+7$
can be obtained using the Jackiw-Nohl-Rebbi (JNR) ansatz \cite{JNR},
where the data consists of $N+1$ distinct points in
$\R^4$ together with a positive weight for each point. 
If all the weights are taken to be equal and the points placed
at the vertices of a polyhedron in $\R^3\subset\R^4$
then the instanton inherits the
symmetries of the polyhedron and the topological charge density is
localised along the edges (and particularly the vertices) of the polyhedron. 
This JNR approach to symmetric instantons therefore provides an upper
bound on the minimal charge required to obtain an instanton associated
with a given platonic solid, that is, $N\le V-1,$ where $V$ is the
number of vertices of the platonic solid. 
For the platonic solids with triangular faces (the tetrahedron, 
octahedron and icosahedron) the minimal instanton charge is equal to
$V-1$ (that is, $N=3,5,11$ respectively) and the instanton is of the 
JNR type. 
However, for instantons associated with the cube and the dodecahedron the 
minimal charges are $4$ and $7$ respectively \cite{LM,SiSu}, which is less than
$V-1,$ as these instantons are obtained using the ADHM construction
and are not of the JNR type. Making use of connections between instantons,
Skyrmions and monopoles, leads to an understanding of this minimal instanton
number as $N=1+\frac{1}{2}F,$ where $F$ is the number of faces of the
polyhedron, and the key ingredient is the fact that
a degree $N$ rational map between Riemann spheres has $2N-2$
ramification points \cite{HMS}.

All the work mentioned above on symmetric instantons 
in $\R^4$ concerns platonic
symmetries, that is, finite subgroups of $SO(3).$ This involves 
breaking the symmetries of $\R^4$ by selecting a distinguished direction, 
leaving an $SO(3)$ rotational symmetry in the remaining
3-dimensional space. 
Motivated by the interesting features found for platonic instantons and 
their subsequent utility, this leads on to a rather natural question
concerning the existence of symmetric instantons in $\R^4,$ where all 
directions are 
treated on an equal footing, so that the full rotational symmetry group 
$SO(4)$ acts. In this case the role of the five platonic solids in three 
dimensions is played by the six regular polytopes in four dimensions,
with the symmetry groups of the pentatope, hyperoctahedron/tesseract, 24-cell
and 120-cell/600-cell, being the relevant finite subgroups of $SO(4).$
In this paper we address this issue by describing the action of these
symmetry groups on the ADHM data of the instanton and providing a framework 
for the explicit construction of this symmetric data. 
 
In analogy to the platonic situation described above, the JNR ansatz can 
be applied to construct instantons with the symmetries of the regular polytopes,
by placing equal weight points at the vertices of the polytope. 
There is therefore again an upper bound,  $N\le V-1,$
on the minimal charge required to obtain an instanton associated
with a given polytope, where $V$ is the number of vertices of the 
polytope. In the platonic case, a prediction for the minimal charge
can be obtained from the properties of rational maps between Riemann spheres,
but there is no similar understanding for polytopes, so it is unknown whether
or not one should expect the minimal charge instantons to saturate this
bound and be of the JNR type. It appears that the only way to answer this 
question is to invoke the ADHM construction and explicitly calculate the
symmetric ADHM data.
 
The double cover of $SO(4)$ is $SU(2)\times SU(2)$ and, 
as we shall see, the pentatope differs from the other polytopes in that
the left and right $SU(2)$ actions are not independent, but rather
there is an action of a twisted diagonal subgroup. This leads to a
simplification that allows the techniques developed for finding platonic ADHM
data to be modified in a simple way to apply to ADHM data with the
symmetries of the pentatope. We describe this modification and apply
an analysis to construct the ADHM data of a charge 4 instanton with
the symmetries of the pentatope. Furthermore, we prove that there
are no instantons of lower charge with this symmetry.
  
For the remaining polytopes, there are independent left and 
right actions of finite subgroups of $SU(2)$ corresponding to 
the binary versions of the dihedral group $D_2$, the tetrahedral
group $T$ and the icosahedral group $Y,$ for the
hyperoctahedron, 24-cell and 120-cell respectively. 
The McKay correspondence \cite{McK} associates these 
three groups to the extended Dynkin diagrams of the 
affine Lie algebras $\widetilde d_4,$ $\ \widetilde e_6,$ and 
$\widetilde e_8.$ This proves to be a useful tool in our work, 
as we note that the building blocks of symmetric ADHM data
are classified by the edges in the Dynkin diagram.  
Using these building blocks we describe a framework in which
they can be assembled into symmetric ADHM data through the identification of 
pairs of commuting representations of the associated binary polyhedral group.
We illustrate our procedure by the construction of ADHM data associated
with the hyperoctahedron and the 24-cell, with instanton
charges 7 and 23, respectively. Furthermore, we show that within our
framework these are the lowest possible charges with these symmetries and
we present plots of topological charge densities that confirm the
polytope structure.

In the three examples of symmetric ADHM data that we have explicitly
constructed, the charge is equal to that given by the JNR upper bound.
We clarify this issue by demonstrating the equivalence of our ADHM data to 
JNR data and make some further comments regarding our current 
understanding of this aspect. 

\section{The regular polytopes and their symmetries}\news \label{sec:regular polytopes}
The platonic solids are regular polyhedra, where all of the faces are identical regular polygons. In four dimensions, the analogue of the platonic solids are the regular polytopes, which are constructed from identical cells that are platonic solids. 
There are several choices of nomenclature for the six regular polytopes, 
including the convention in which each is named after the number of 
3-dimensional cells it contains:
\begin{itemize}
\item The pentatope, or 5-cell, is the 4-dimensional analogue
of the tetrahedron and is self-dual. 
It consists of 5 vertices, 10 edges and 10 triangular faces forming 5 tetrahedra.
 \item The tesseract, or 8-cell, is the 4-dimensional 
analogue of the cube and is dual to the hyperoctahedron.
It consists of 16 vertices, 32 edges and 24 square faces forming 8 cubes.
\item The hyperoctahedron, or 16-cell, is the 4-dimensional analogue of the octahedron and is dual to the tesseract.
It consists of 8 vertices, 24 edges and 32 triangular faces forming 16 tetrahedra.
\item The octaplex, or 24-cell, is self-dual and is unique to four dimensions,
having no analogue in any other dimension.
It consists of 24 vertices, 96 edges and 96 triangular faces forming 24 octahedra.
\item The dodecaplex, or 120-cell, is the 4-dimensional \edit{analogue} of the dodecahedron and is dual to the tetraplex.
It consists of 600 vertices, 1200 edges and 720 pentagonal
 faces forming 120 dodecahedra.
\item The tetraplex, or 600-cell, is the 4-dimensional analogue of the icosahedron and is dual to the dodecaplex.
It consists of 120 vertices, 720 edges and 1200 triangular faces forming 600 tetrahedra.
\end{itemize}

The regular polytopes that are dual to each other share the same symmetry group, 
so the only symmetry groups that we need to consider are that of the 5-cell, 
the 16-cell, the 24-cell and the 600-cell. 
As we shall see, the explicit implementation of our general framework is stretched to its limit with the 24-cell, so we shall only briefly mention the 
application to the 120-cell and 600-cell in this paper. In the rest of this section we shall review the above symmetry groups,
following \cite{Du}.

If $\R^4$ is identified with the quaternions, then the action of any rotation, $\g \in \SO(4)$, on $x\in\R^4,$ can be expressed as left and right multiplication by unit quaternions,
\begin{equation}
\g \circ x = g_L \, x \, g_R^{-1},
\label{leftandright}
\end{equation}
for some unit quaternions $g_L$ and $g_R,$ which may be identified
with elements of $SU(2)$. The action of $(g_L, g_R)$ is identical to the action of $(-g_L, -g_R)$, so there are two elements in $\SU(2) \times \SU(2)$ which correspond to the same element in $\SO(4)$, reflecting the fact
that $\SU(2) \times \SU(2)$ is the double cover of $\SO(4)$. 
The symmetry groups of the 5-, 16-, 24- and 600-cell are all naturally expressed as subgroups of $\SU(2) \times \SU(2)$, with the true symmetry group being the projection to $\SO(4)$.

\subsection{The 5-cell}\label{the5cell}
The symmetry group of the 5-cell is realised in a different way to 
that of the other polytopes, so we shall consider it first.

The binary icosahedral group $\BY,$ is generated by the unit quaternions
\begin{equation}
g_1 = \tfrac{1}{2}(1 + i + j+ k), \quad \text{and} \quad g_2 = \tfrac{1}{2}(\tau + \tau^{-1} i + j),
\label{geny}
\end{equation}
where $\tau = \tfrac{1}{2} \big( \sqrt{5} + 1 \big).$
These generators satisfy the relations
\begin{equation}
g_1^\alpha = g_2^\beta = (g_1 g_2)^\gamma = -1, \quad \text{with } \alpha = 3,\, \beta = 5,\, \gamma = 2.
\end{equation}
The vertices of the 5-cell can be taken to be the five unit quaternions,
\begin{equation}
1, \tfrac{1}{4} \del{ -1 \pm i \pm j \pm k },
\end{equation}
where an odd number of plus signs is taken for each vertex. 
These vertices are permuted under the action of $\BY,$
through a twisted diagonal embedding into $SU(2)\times SU(2).$
Explicitly,
\begin{equation}
x \mapsto g^\sharp x g^{-1},
\label{twistedleftandright}
\end{equation}
where $g\in \BY$, 
and $g^\sharp$ is the dual of $g$, obtained by making the replacement 
$\sqrt{5}\mapsto -\sqrt{5}$ in the generators. 
The double cover of the symmetry group of the 5-cell is therefore a subgroup of $\SU(2)$ that is embedded in $\SU(2) \times \SU(2)$ via $g \mapsto ( g^\sharp, g)$.

\subsection{The 16-cell}\label{sub16}
The 8 vertices of the 16-cell may be taken to be at the intersection points
of the four Cartesian axes with the unit four-sphere. As quaternions 
these vertices form the binary dihedral group $\BD_2$,
also known as the quaternion group,
\begin{equation}
\Q = \{ \pm 1, \pm i, \pm j, \pm k \}.
\end{equation}
$\Q$ is a group under quaternionic multiplication and the left and right action of $\Q$ permutes the vertices of the $16$-cell
\begin{equation}
x \mapsto g_L \, x \, g_R^{-1}, \quad\quad g_L, g_R \in \Q.
\end{equation}
The double cover of the rotational symmetry group of the 16-cell is 
$\Q \times \Q \subset \SU(2) \times \SU(2)$, 
where $\Q$ is generated by the two elements
\begin{equation}
g_1 = i, \quad \text{and} \quad g_2 = j.
\end{equation}
The quaternion group generators satisfy
\begin{equation}
g_1^\alpha = g_2^\beta = (g_1 g_2)^\gamma = -1, \quad \text{with } \alpha = \beta = \gamma = 2.
\end{equation}
The 16-cell is dual to the 8-cell, which shares the same symmetry group.

\subsection{The 24-cell}\label{sub24}
The symmetry of the 24-cell has a similar structure to that 
of the 16-cell, with the binary dihedral group $\Q$ replaced
by the binary tetrahedral group $\BT.$
The 24 vertices may be taken to be
\begin{equation} \label{eq:vertices of 24-cell}
\BT = \{ \pm 1, \pm i, \pm j, \pm k, \tfrac{1}{2}\del{ \pm 1 \pm i \pm j \pm k} \}. 
\end{equation}
which forms the group $\BT$ under multiplication.
The double cover of the rotational symmetry group of the 24-cell is $\BT \times \BT$, with rotations acting via left and right multiplication. 
$\BT$
is generated by
\begin{equation}
g_1 = \tfrac{1}{2}(1 + i + j+ k), \quad \text{and} \quad g_2 = \tfrac{1}{2}(1 + i+ j -k),
\label{genbt}
\end{equation}
which satisfy
\begin{equation}
g_1^\alpha = g_2^\beta = (g_1 g_2)^\gamma = -1, \quad \text{with } \alpha = \beta = 3,\, \gamma = 2.
\end{equation}

\subsection{The 600-cell}
In a similar fashion to the 16-cell and 24-cell, the 120 vertices of the 
600-cell form the binary icosahedral group 
\begin{equation}
\BI = \{ \pm 1, \pm i, \pm j, \pm k, \tfrac{1}{2}\del{ \pm 1 \pm i \pm j \pm k}, \tfrac{1}{2} \del{\pm i \pm \tau j \pm \tau^{-1} k } \}. 
\end{equation}
The double cover of the rotational symmetry group of the 600-cell is therefore $\BI \times \BI$, with rotations acting via left and right multiplication. 
The generators of $\BY$ have already been presented in (\ref{geny}). 
The 600-cell is dual to the 120-cell, which shares the same symmetry group.

\subsection{Group representations and the McKay correspondence}
\label{McKay}
An $n$-dimensional representation of a group is a map, 
$\rho$, from the group to $\GL(n)$, such that the matrices, $\rho(g)$, preserve the group relations. 
For the binary polyhedral groups of interest in this paper, 
the representation matrices must satisfy the group relations
\begin{equation} \label{eq:generators of Q8}
\rho(g_1)^\alpha = \rho(g_2)^\beta = \del{ \rho(g_1) \rho(g_2) }^\gamma.
\end{equation}
In our application to ADHM data we shall mainly be concerned with 
real representations, so it will be important to identify real 
irreducible representations, taking into account that
representations that are reducible over $\C$ may be irreducible over $\R.$ 

All irreducible $n$-dimensional representations of the binary polyhedral
group satisfy
\begin{equation}
\rho(g_1)^\alpha = \rho(g_2)^\beta = \del{ \rho(g_1) \rho(g_2) }^\gamma = \varepsilon\id_n, \quad \text{where $\varepsilon = \pm 1$}.
\end{equation}
If $\epsilon=1$ then the representation is also a representation of the
polyhedral group, that is, of the finite subgroup of $SO(3),$ and we shall
refer to this as a positive representation.
If $\epsilon=-1$ then the representation is not a representation of the
polyhedral group but only of the binary polyhedral group, that is, of the 
finite subgroup of $SU(2),$ and we shall
refer to this as a negative representation.

Our nomenclature for irreducible representations follows the notation
that is common in chemistry, where representations are labelled by a letter 
which indicates their dimension. 
1-dimensional representations are labelled by $A$, while 2-dimensional representations are labelled by $E$, 3-dimensional representations by $F$, and higher dimensions by going through the alphabet in sequence. Negative representations are indicated by a prime, for example, 
$G'$ denotes a 4-dimensional negative representation. 
The fundamental quaternion representation, when viewed as a complex representation, is 2-dimensional and we shall denote it by $E'$. 
It is the fundamental representation obtained by the
restriction of the 2-dimensional irreducible representation of $SU(2)$
to the binary polyhedral group

The McKay correspondence \cite{McK} provides a mapping between 
the 
binary polyhedral groups 
and the Dynkin diagrams of the affine
simply-laced Lie algebras. For the binary polyhedral groups of interest
in this paper, $\Q,\BT,\BI,$ the associated affine Lie algebras 
are $\widetilde d_4,$ $\, \widetilde e_6,$ and $\widetilde e_8,$ respectively. 
There is a one-to-one correspondence between the irreducible representations
of the binary polyhedral group and the nodes in the extended Dynkin 
diagram. Furthermore, the nodes associated with the representations
$\rho_i$ and $\rho_j$ are joined by an edge if and only if $\rho_j$ is
contained in the decomposition of $\rho_i\otimes E',$ where 
$E'$ denotes the fundamental 2-dimensional representation, as described
above. These features will prove to be useful in our computations.

\section{Group actions on the ADHM data}\news \label{sec:group actions on the ADHM data}
In this section we shall discuss symmetric instantons
and describe how the symmetry group acts on both the instanton and the 
underlying ADHM data. For an instanton to be symmetric, the 
gauge potential after the action of the symmetry 
must be gauge equivalent to the original gauge potential.
As a consequence of this, the topological charge density of the instanton 
is invariant under the action of the symmetry group. 
For the moment, the symmetry can be any subgroup of $\SO(4)$, 
such as the symmetry groups of the platonic solids, or of the regular 
polytopes. 
The gauge potential of an instanton, $a_i(x)\in su(2)$ for
$i=1,\ldots,4$
is associated to its ADHM data, from which it 
can be constructed. In this section we shall see how the action of the 
symmetry group can be lifted to an action on the ADHM data.

If $\G \subset \SO(4)$ is the symmetry group of an instanton then
for each $\g\in\G$ there must exist $\Omega_\g(x)\in SU(2)$ such that
\begin{equation}
a_i(\g \circ x) = \Omega_\g(x) a_i(x) \Omega_\g^{-1}(x) + \Omega_\g(x) \partial_i ( \Omega_\g^{-1}(x) ).
\end{equation}

To find solutions with this symmetry we need to lift the action of 
$\G$ on $a_i$ to an action on the underlying ADHM data. 
Recall that the ADHM data for a charge $N$ instanton with 
gauge group $SU(2)$ is given by
\begin{equation}
\Delta(x) = \M - U x,
\end{equation}
where
\begin{equation}
\M = \begin{pmatrix}
L \\
M
\end{pmatrix},
\quad \text{and} \quad
U = \begin{pmatrix}
0 \\
\id_N
\end{pmatrix}.
\label{standardform}
\end{equation}
In this expression $L$ is a length $N$ quaternionic row vector and 
$M$ is an $N \times N$ symmetric quaternionic matrix, which together
satisfy the ADHM constraint that
$\M^\dagger\M$ is a real non-singular matrix, where $\dagger$ denotes
the quaternionic conjugate transpose.  
The spatial coordinate, $x$, is a quaternion in this construction,
where $\R^4$ is identified with the quaternions as described earlier.

The gauge potential is obtained in terms of
an $(N+1)$-component column vector, $\Psi$, of unit length $\Psi^\dagger \Psi = 1$, 
that solves the linear problem
\begin{equation}
\Psi^\dagger \Delta = 0. \qquad 
\end{equation}
The explicit formula for the gauge potential is
\begin{equation}
a_i = \Psi^\dagger \partial_i \Psi,
\end{equation}
where a pure quaternion (that is, with no real component)
 is identified with an element of $su(2).$
Note that $\Psi$ is unique only up to right multiplication by a unit 
quaternion, and this corresponds to a gauge transformation of $a_i$. 

The topological charge density ${\cal N},$ whose integral over
$\R^4$ gives the topological charge $N,$ is given by
\begin{equation}
{\cal N}=-\frac{1}{32 \pi ^2}\,\varepsilon_{ijkl} \Tr \del{ f_{ij} f_{kl} },
\label{chargedensity}
\end{equation}
where $f_{ij}$ is the gauge field 
$f_{ij}=\partial_i a_j-\partial_j a_i+[a_i,a_j].$
 This has the following useful expression in terms of the ADHM data 
\begin{equation}
f_{ij} = - \Psi^\dagger U (\Delta^\dagger \Delta)^{-1} (e_i \bar e_j - e_j \bar e_i) U^\dagger \Psi,
\label{chargefromadhm}
\end{equation}
where $e_i = \{i,j,k,1\}$.

If the gauge potential is symmetric under the action of $\G$ then the 
ADHM data, $\Delta(x)$, must also transform in a way that 
yields a gauge equivalent gauge potential.
The required transformation of the ADHM data takes the form
\begin{equation} \label{eq:symmetry condition on ADHM data}
\Delta \rightarrow
\begin{pmatrix}
p & 0 \\
0 & P
\end{pmatrix} \Delta R^{-1},
\end{equation}
where $p$ is a unit quaternion, $P$ is an $N \times N$ quaternionic matrix 
such that $P^\dagger P = \id_N$, and $R$ is an invertible $N \times N$ 
quaternionic matrix. For each element of the symmetry group, $\g \in \G$, 
the transformed ADHM data is $\Delta(\g \circ x)$. 
For a symmetric instanton, this must be equivalent to $\Delta(x)$, so there 
exists $p_\g$, $P_\g$ and $R_\g$ such that
\begin{equation}
\Delta(\g \circ x) = \begin{pmatrix}
p_\g & 0 \\
0 & P_\g
\end{pmatrix} \Delta(x) R_\g^{-1}.
\end{equation}
Recall that every rotation in $\R^4$ can be represented by left and right multiplication by unit quaternions,
\begin{equation}
\g \circ x = g_L \, x \, g_R^{-1}.
\end{equation}
Since the ADHM data is quaternionic, this is a natural way to represent the action of $\g$. By comparing the terms in \eqref{eq:symmetry condition on ADHM data} that are linear in $x$, we see that $P_\g$ and $R_\g$ must factor into 
\begin{equation}
P_\g = Q_\g \, g_L, \quad \text{and} \quad R_\g = Q_\g \, g_R,
\end{equation}
for some real orthogonal matrix, $Q_\g$. The left quaternion, $g_L$, may also be factored out of $p_\g$, so that for symmetric ADHM data there must exist a quaternion $q_\g$, and a real orthogonal matrix, $Q_\g$, such that
\begin{equation}
\Delta(g_L \, x \, g_R^{-1}) = \begin{pmatrix}
q_\g & 0 \\
0 & Q_\g
\end{pmatrix} g_L \, \Delta(x) \, g_R^{-1} \, Q_\g^{-1}.
\end{equation}
In terms of the blocks in the ADHM data, $L$ and $M$, this condition is
\begin{equation}
Q_\g \, g_L \, M = M \, g_R \, Q_\g \quad \text{and} \quad  
q_\g \, g_L \, L  = L\, g_R \, Q_\g. 
\label{landm}
\end{equation}

To recap, if we have a symmetric instanton, then its ADHM data must be 
invariant under the action of each symmetry, $\g \in \G$. This can be 
represented by the action of an element in the double cover of $\G$, 
which is a subgroup of $SU(2) \times SU(2)$, and acts by left and right 
quaternion multiplication. The transformed ADHM data must give a gauge
equivalent gauge potential, and so there must exist $q_\g$ and $Q_\g$, 
as above, that relate it back to the original ADHM data. 
The construction of symmetric ADHM data involves using representation
theory to determine possible choices for $q_\g$ and $Q_\g,$ which then
produces a simplified form for $L$ and $M,$ to which the ADHM constraint
can be applied.

\section{The ADHM 5-cell}\news \label{sec:5-cell}
As discussed above, the action of the 5-cell symmetry group is different 
from the other polytopes because the left and right actions do not act 
independently. This allows the existing machinery developed to study
platonic instantons to be modified in a simple way to apply
to this situation, as follows.
We recall that the instantons considered 
in \cite{SiSu} are symmetric under
the icosahedral group $Y\subset SO(3)$. 
The binary icosahedral group $\BY \subset SU(2)$, 
acts by quaternionic multiplication, as given by 
(\ref{leftandright}), with the diagonal embedding of
$SU(2)$ into $SU(2)\times SU(2)$ given by $(g_L,g_R)=(g,g)$ for
$g\in\BI.$

As described in Section \ref{the5cell}, for the 5-cell
 this action is twisted, with the left quaternion replaced by the dual, 
$g^\sharp$, to give the twisted diagonal embedding
$(g_L,g_R)=(g^\sharp,g)$ with $g\in\BI.$
The representation theory remains largely the same, though the twisting
turns out to allow a symmetric instanton with a lower charge than in
the untwisted case, as we now show.

Given the above discussion, in (\ref{landm}) we take $g_R=g\in \BY$ with
$g_L=g^\sharp$ and $Q_\g=Q(g).$ The matrices $Q(g)$ form a real 
$N$-dimensional representation of $\BY,$ and with a slight abuse of notation
we shall use $Q$ to denote both the abstract representation and the explicit
matrices. For the symmetry of the 5-cell it turns out to be more
convenient not to factor out the left quaternion from 
$p_\g,$ so we write $p_\g=q_\g g^\sharp=p(g).$ The quaternions $p(g)$ form
a representation of $\BY$ that may be viewed as a complex 2-dimensional
representation, which we also denote by $p.$ 
The condition for ADHM data $L,M$ to have the symmetry of
the 5-cell therefore becomes
\begin{equation}
Q(g_i) \, g_i^\sharp \, M = M \, g_i \, Q(g_i) \quad \text{and} \quad  
p(g_i)\, L  = L\, g_i \, Q(g_i) 
\label{landm5cell}
\end{equation}
where $i=1,2$ and $g_1,g_2$ are the generators of $\BY$ given in
(\ref{geny}).

\begin{figure}
\centering
\includegraphics[width=16.5cm]{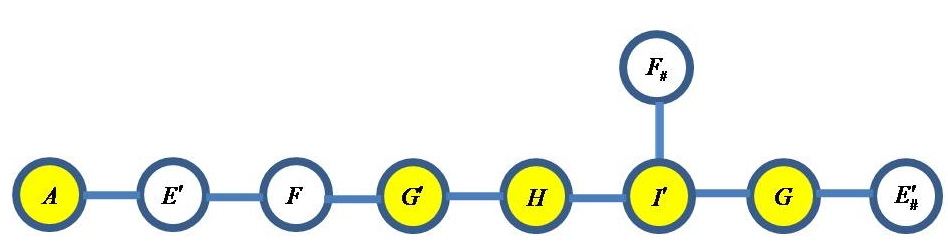}
\caption{
The Dynkin diagram of $\widetilde e_8$ providing a graphical
illustration of the irreducible representations of $\BY$ via the
McKay correspondence. The filled nodes denote 
self-dual representations.}
\label{fig:e8}
\end{figure}

The representation theory of $\BY$ is captured by the Dynkin diagram
of $\widetilde e_8,$ presented in Figure~\ref{fig:e8}.
There are nine irreducible representations of $\BY,$ with one
for every dimension from one up to six  obtained as the restriction
of the corresponding irreducible representation of $SU(2).$ 
Using the notation discussed in Section \ref{McKay}, 
we denote these six representations by
$A,E',F,G',H,I'.$
As for the three remaining representations, 
$E'_\sharp$ is the dual 2-dimensional representation obtained from
the representation $E'$ by making the replacement 
$\sqrt{5}\rightarrow -\sqrt{5}$ in the character table.
Similarly, there is a 3-dimensional representation,
$F_\sharp,$ that is dual to $F.$ The final representation is
the 4-dimensional representation $G=E'\otimes E'_\sharp.$  
The representations $E,E_\sharp$ and $F,F_\sharp$ are dual pairs
and we shall use the term self-dual for all the other 
irreducible representations, which we indicate by filled
nodes in the Dynkin diagram.  

As $Q$ is a real $N$-dimensional representation then it must have a 
decomposition
into the irreducible real representations $A,F,F_\sharp,G,H,$
as the remaining irreducible representations are complex.
Furthermore, as the 5-cell has 5 vertices then the JNR bound for the minimal 
charge is $N\le 4,$ so with this restriction the 5-dimensional representation
$H$ is already ruled out.
We can neglect the trivial representation, so the only possibilities 
for $Q$ that remain to be investigated are $F,F_\sharp$ and $G.$

From (\ref{landm5cell}) we see that $M$ is an invariant map
\be M:\ Q\otimes E'\mapsto Q\otimes E_\sharp'\ee
and 
$L$ is an invariant map
\be L:\ Q\otimes E'\mapsto p.\ee

By the McKay correspondence, the invariance of $L$ requires that 
in the Dynkin diagram of $\widetilde e_8$
the node associated to the representation $Q$ must be joined by an
edge to the node associated with the 2-dimensional representation $p.$
This eliminates the possibility that $Q$ is equal to $F_\sharp,$ since this node 
is not joined to the node of any 2-dimensional representation.

As $Q\otimes E'$ is equal to the nodes joined to $Q,$ then
taking the dual of this relation we see that
$Q\otimes E_\sharp'$ is equal to the dual of the nodes joined to
$Q_\sharp,$ where $Q_\sharp=Q$ if $Q$ is self-dual.
The invariance of $M$ therefore requires that there is a node common to the
nodes joined to $Q$ and the dual of the nodes joined to $Q_\sharp.$
This rules out the possibility that $Q=F,$ since the nodes joined to
$F$ are $E'$ and $G',$ whereas the dual of the only node joined to $F_\sharp$ is
$I'.$  

The only remaining possibility is $Q=G,$ and this does yield an invariant map.
In this case the nodes joined to $Q$ are $I'$ and $E_\sharp'$ and the
dual of the nodes joined to $Q_\sharp=Q$ are $I'$ and $E'.$ The node 
$I'$ is common to both sets and therefore there is an 
associated invariant map $M.$
As $G$ is joined to the node $E_\sharp'$ then there is an invariant map $L$ 
with $p=E_\sharp'.$ 

With this choice, 
and using the canonical basis for $G$ in which
\begin{equation}
\begin{gathered}
G(g_1) = \begin{pmatrix}
1 & 0 & 0 & 0 \\
0 & 0 & 0 & 1 \\
0 & 1 & 0 & 0 \\
0 & 0 & 1 & 0
\end{pmatrix}, \quad\quad
G(g_2) = \frac{1}{4} \begin{pmatrix}
-1 & \sqrt 5 & -\sqrt 5 & \sqrt 5 \\
-\sqrt 5 & -3 & -1 & 1 \\
\sqrt 5 & -1 & 1 & 3 \\
\sqrt 5 & -1 & -3 & -1
\end{pmatrix},
\end{gathered}
\end{equation}
equations (\ref{landm5cell}) become
\begin{equation}
G(g_i) \, g_i^\sharp \, M = M \, g_i \, G(g_i) \quad \text{and} \quad  
g_i^\sharp\, L  = L\, g_i \, G(g_i). 
\label{landm5cellagain}
\end{equation}
Solving these linear equations for $L$ and $M$ yields
\be
L=l_0(1, i, j, k),
 \quad \mbox{and} \quad\
M = b_0\begin{pmatrix}
-3 & i & j & k \\
i & 1 & - \sqrt 5 k & - \sqrt 5 j \\
j & - \sqrt 5 k & 1 & -\sqrt 5 i \\
k & -\sqrt 5 j & -\sqrt 5 i & 1
\end{pmatrix},
\ee
where $l_0$ and $b_0$ are arbitrary real parameters.

Imposing the ADHM constraint on this data reduces to the requirement
that $l_0^2=4b_0^2$ and without loss of generality we can choose
$b_0=\lambda=-\frac{1}{2}l_0$ to give the ADHM data.
\begin{equation} \label{5cellADHM}
\M = \lambda \begin{pmatrix}
-2 & -2i & -2j & -2k \\
-3 & i & j & k \\
i & 1 & - \sqrt 5 k & - \sqrt 5 j \\
j & - \sqrt 5 k & 1 & -\sqrt 5 i \\
k & -\sqrt 5 j & -\sqrt 5 i & 1
\end{pmatrix},
\end{equation}
where $\lambda$ is a real parameter that determines the scale of the instanton. 

\begin{figure}
\centering
\includegraphics[width=5cm]{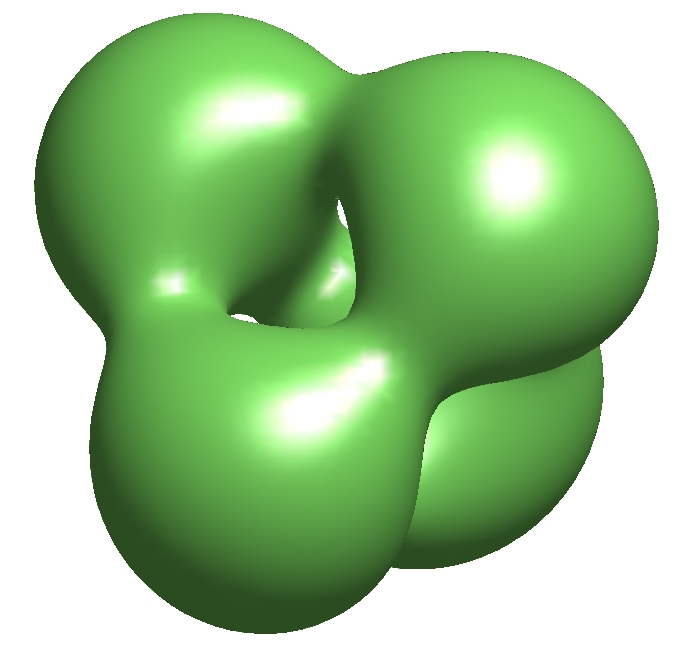}
\includegraphics[width=5cm]{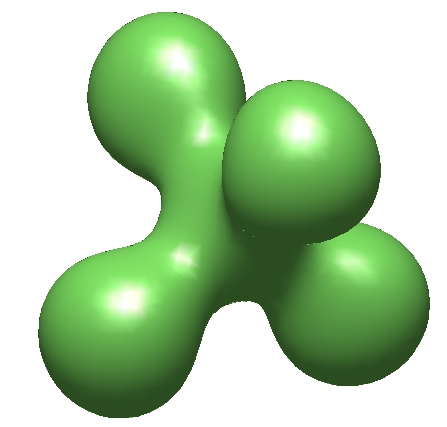}
\caption{
Surfaces of constant topological charge density for the
charge $4$ instanton with the symmetries of the $5$-cell. 
The left image is the charge density in the hyperplane
$x_4=0,$ where the vertices of the $5$-cell form a tetrahedron.
The right image is the charge density integrated along the $x_4$-direction. 
}
\label{fig:topological charge density of 5-cell instanton}
\end{figure}

\edit{The vertices of the 5-cell that lie in the $x_4=0$ hyperplane form a
tetrahedron.
The topological charge density (\ref{chargedensity}) of the 5-cell 
instanton in this hyperplane is plotted as an isosurface in the left image in 
Figure \ref{fig:topological charge density of 5-cell instanton},
using the formula (\ref{chargefromadhm}). The tetrahedron is clearly
visible in this image. The right image in  
Figure \ref{fig:topological charge density of 5-cell instanton}
captures more of the information about all the vertices by displaying
an isosurface of the charge density integrated along the $x_4$-direction.}

We have shown that this charge 4 ADHM data is invariant under the action 
of the symmetry group of the 5-cell, and moreover that there is no instanton
of lower charge with this symmetry. As this charge is equal to the
JNR bound then this ADHM data must be equivalent to JNR data in which the
five points are placed at the vertices of a 5-cell with equal weights.
We prove this equivalence explicitly in Section \ref{sec:JNR}.    

\section{\edit{Constructing ADHM polytopes}}
\news \label{sec:left and right screws}
Consider ADHM data that is invariant under the symmetries of 
one of the polytopes other than the 5-cell.
The crucial difference is that now the left and right actions 
can be applied independently. First consider the right action of the generators of the binary polyhedral group, so that $g_L=1$ and $g_R \in\{g_1, g_2\}$. 
Then there must exist matrices $Q_{R}(g_i)$ and quaternions $q_R(g_i),$ 
for $i=1,2,$ which satisfy
\begin{equation}
Q_{R}(g_i) \, M  = M g_i \,Q_{R}(g_i) \quad \text{and} \quad q_R(g_i) L = L \, g_i \, Q_{R}(g_i).
\label{rightaction}
\end{equation}
In our framework, these matrices in the right action form a 
real $N$-dimensional representation of the binary polyhedral group,
\begin{equation}
(Q_{R}(g_1))^\alpha = (Q_{R}(g_2))^\beta = \del{ Q_{R}(g_1) Q_{R}(g_2) }^\gamma.
\end{equation}
In keeping with the earlier notation, we shall use $Q_R$ to denote both the
abstract representation and the associated explicit matrices, given a basis 
for the representation that will be specified later.
As we are free to choose a basis for $Q_R,$ we can decompose it into the 
direct sum of irreducible representations. 
If we order the irreducible representations so that the positive representations form the upper blocks of $Q_R$ and the negative representations form the lower blocks of $Q_R$ then
\begin{equation}
(Q_{R}(g_1))^\alpha = (Q_{R}(g_2))^\beta = \del{ Q_{R}(g_1) Q_{R}(g_2) }^\gamma = \begin{pmatrix}
\id_m & 0 \\
0 & -\id_n
\end{pmatrix},
\end{equation}
where $m+n = N$. We write $Q_R=Q_R^+\oplus Q_R^-$ as
\begin{equation}
Q_R(g) = \begin{pmatrix}
Q_R^+(g) & 0 \\
0 & Q_R^-(g)
\end{pmatrix},
\end{equation}
where $Q_R^+$ is an $m$-dimensional positive representation and $Q_R^-$ 
is an $n$-dimensional negative representation of the binary polyhedral group.

From (\ref{rightaction}) we see that $M$ \edit{is} an invariant map from 
the $N$-dimensional representation $Q_R$ tensored with the quaternion 
representation $E'$, 
back to the representation $Q_R$
\begin{equation}
M : Q_R \otimes E' \mapsto Q_R.
\end{equation}
As $E'$ is a negative representation then 
$Q_R^+\otimes E'$ can have no component in common with $Q_R^+$
and    
$Q_R^-\otimes E'$ can have no component in common with $Q_R^-.$ 
Thus $M$ must map the component of a negative representation in $Q_R^-$ to 
the component of a positive representation in $Q_R^+$, or vice-versa. 
As $M$ is a symmetric matrix, then in our given basis it must take the
off-diagonal form
\begin{equation}
M = \begin{pmatrix}
0 & B \\
B^\T & 0
\end{pmatrix},
\end{equation}
where $B$ is an $m \times n$ quaternionic matrix.

Consider the irreducible 
(over $\R$) decomposition 
$Q_R=Q_{R,1}^+\oplus\cdots\oplus Q_{R,s}^+\oplus
Q_{R,1}^-\oplus\cdots\oplus Q_{R,t}^-$
with the implied block structure
\begin{equation}
Q_R = \begin{pmatrix}
\begin{matrix}
Q_{R,1}^+ & & \\
& \ddots & \\
& & Q_{R,s}^+
\end{matrix} & \\
& \begin{matrix}
Q_{R,1}^- & & \\
& \ddots & \\
& & Q_{R,t}^-
\end{matrix}
\end{pmatrix}.
\end{equation}
The ADHM data can also be decomposed into this block form, in particular
\begin{equation}
M = \begin{pmatrix}
0 & \begin{matrix} B_{11} & \cdots & B_{1t} \\ \vdots & \ddots & \vdots \\
B_{s1} & \cdots & B_{st}
\end{matrix} \\
\begin{matrix} B_{11}^\T & \cdots & B_{s1}^\T \\ \vdots & \ddots & \vdots \\
B_{1t}^\T & \cdots & B_{st}^\T
\end{matrix} & 0
\end{pmatrix},
\end{equation}
where the block $B_{uv}$ satisfies
\begin{equation} 
Q_{R,u}^+(g_i) \, B_{uv} = B_{uv} \, g_i \, Q_{R,v}^-(g_i),
\label{Brightaction} 
\end{equation}
with $u = 1, \ldots s$, and $v = 1, \ldots t$, 
which are not summed over in the above expressions.
From this equation we see that $B_{uv}$ is an invariant map 
\be
B_{uv}: Q_{R,v}^-\otimes E'\mapsto Q_{R,u}^+,
\ee
and therefore exists if and only if the representation $Q_{R,u}^+$ 
is contained
in the irreducible decomposition of $Q_{R,v}^-\otimes E'.$ 
However, we observe that this is
precisely the condition that the nodes associated with the
representations $Q_{R,u}^+$ and $Q_{R,v}^-$ are joined by an edge in
the Dynkin diagram obtained using the McKay correspondence. In this way
the building blocks of symmetric ADHM data are labelled by the edges in
the Dynkin diagram. To complete this description we must add an extra label
to the nodes that correspond to complex representations,
since $Q_R$ is a real representation. Explicitly, we introduce the notation
$\rho_j[\rho_k]$ to denote that $\rho_j$ is a complex irreducible representation
and $\rho_k$ is a real representation that is irreducible over $\R$ but
is reducible over $\C$ and contains $\rho_j$ in its decomposition 
into irreducible components. Each edge in the Dynkin diagram now
corresponds to an invariant block between real representations, where
we associate the real representation $\rho_k$ with the node
$\rho_j[\rho_k]$ and note that two edges may now be associated with the
same invariant block, since two different complex representations may
appear in the decomposition of the same real representation.

For each edge in the Dynkin diagram the associated invariant map 
given by the matrix $B_{uv}$, 
of size $\mbox{dim}(Q_{R,u}^+)\times\mbox{dim}(Q_{R,v}^-),$ can be
obtained explicitly by solving the linear equation (\ref{Brightaction}).
This matrix will contain free parameters, for example it is clear from
(\ref{Brightaction}) that there is the freedom to multiply $B_{uv}$
on the left by an arbitrary quaternion. In what follows it will be convenient to
treat multiple copies of the same irreducible representation as a single representation,
with the invariant map constructed from the single invariant block by the obvious
tensor product.

If we now consider the left action, 
$g_R=1$ and $g_L \in\{g_1, g_2\}$, 
then there must exist matrices $Q_{L}(g_i)$ and quaternions $q_L(g_i)$ which satisfy
\begin{equation}
Q_{L}(g_i) \, g_i \, M  = M \,Q_{L}(g_i) \quad \text{and} \quad q_L(g_i) \, g_i \, L = L \, Q_{L}(g_i).
\label{leftaction}
\end{equation}
The representation of the left action can be put in the same block form as the right action,
\begin{equation}
Q_L(g) = \begin{pmatrix}
Q_L^+(g) & 0 \\
0 & Q_L^-(g)
\end{pmatrix},
\end{equation}
where $Q_L^+$ is an $m$-dimensional positive representation and $Q_L^-$ is a 
$n$-dimensional negative representation. Furthermore, 
the representation of the left action shares the same block structure
as the right action, 
\begin{equation}
Q_L = \begin{pmatrix}
\begin{matrix}
Q_{L,1}^+ & & \\
& \ddots & \\
& & Q_{L,s}^+
\end{matrix} & \\
& \begin{matrix}
Q_{L,1}^- & & \\
& \ddots & \\
& & Q_{L,t}^-
\end{matrix}
\end{pmatrix},
\end{equation}
where each pair of blocks in the left and right actions, 
$(Q_{L,u}^+, Q_{R,u}^+)$, 
or $(Q_{L,{v}}^-, Q_{R,{v}}^-)$, corresponds to a pair of left and right representations 
of the same dimension, but not necessarily the same representations.
In particular, unlike the right action, these blocks 
in the left action are generally not in a basis where they are 
manifestly the direct sum of irreducible representations, since we no longer have the 
freedom to arbitrarily choose the basis, having already fixed $Q_R$ 
in canonical form as
a direct sum of irreducible representations. This is an important point and
in particular it means that when we refer to a representation $Q_{L,u}^+$ or 
$Q_{L,v}^-$ we need to refer to a specific basis, and two representations which
would usually be considered equivalent will need to be distinguished because 
the change of basis required to map one representation to the other is not 
compatible with preserving the canonical basis for the right representation. 

So far we have considered the left and right actions independently, but the full set of rotations in $\SO(4)$ are generated by acting with both a left and right action together. 
In terms of the action on the spatial coordinate $x\mapsto g_i x g_j^{-1},$
the order of the action is irrelevant, so either the left or right action can be
applied first. 
This implies that the matrices in the left and right representations must commute,
\begin{equation}
Q_{R}(g_i) \, Q_L(g_j) = Q_{L}(g_j) \, Q_{R}(g_i).
\end{equation}
At first sight it may appear that anti-commuting is also a possibility, but this
can be ruled out by considering the left or right action of $(g_1)^{2\alpha} = 1,$ under 
which $(Q_{L,R}(g_i))^{2\alpha} = \id_N$ and so must commute. 

The procedure for constructing symmetric ADHM data will therefore involve 
calculating the invariant building blocks associated with the edges in the
Dynkin diagram and assembling these into invariant data by 
identifying pairs of commuting representations. 
For an appropriate range of charges, such invariant data 
exists and has only a few free parameters, which can then be constrained 
by the ADHM condition to determine whether or not an associated symmetric 
instanton exists.  

The final issue we need to address in this section is that the upper 
row vector in the ADHM data, $L$, 
must also be invariant under the above left and right actions,
as specified by the second equations in both (\ref{rightaction}) and (\ref{leftaction}).
The first of these equations implies that $L$ is an invariant map
\be L:Q_R\otimes E'\mapsto q_R. \ee
We can write $L$ in block form with the same block structure as $Q_R$ and $Q_L$,
\begin{equation}
L = (L^+\; L^-).
\label{lblock}
\end{equation}
With this block structure, $L^\pm$ are invariant maps
\be L^\pm:Q_R^\pm\otimes E'\mapsto q_R. \ee
As $q_R$ is a 1-dimensional quaternionic representation then it is either
positive, in which case $L^+$ must vanish, or 
negative, in which case $L^-$ must vanish. 

A similar consideration 
of the left action shows that $q_L$ must be a representation of the same sign 
as $q_R$ to be able to leave the remaining non-zero block in $L$ invariant. 
By a similar argument to the one given above for the real
representations $Q_L$ and $Q_R$, the quaternion representations
$q_L$ and $q_R$ must commute. However, there are no commuting
negative representations for $q_R$ and $q_L,$ so they must both be
positive representations. 

In summary, $L$ has the block form (\ref{lblock}) with $L^+=0$ and
$L^-$ an invariant map 
\be L^-:Q_R^-\otimes E'\mapsto q_R,\ee
where $q_R$ is a positive representation, and similarly for the left
action. Putting everything together, the complete ADHM data takes the block
form
\begin{equation}
\M = \begin{pmatrix}
0 & \begin{matrix} L^-_1 & \cdots & L^-_t \end{matrix} \\
0 & \begin{matrix} B_{11} & \cdots & B_{1t} \\ \vdots & \ddots & \vdots \\
B_{s1} & \cdots & B_{st}
\end{matrix} \\
\begin{matrix} B_{11}^\T & \cdots & B_{s1}^\T \\ \vdots & \ddots & \vdots \\
B_{1t}^\T & \cdots & B_{st}^\T
\end{matrix} & 0
\end{pmatrix}.
\end{equation}

In this section we have introduced a framework for the construction of
ADHM data with the symmetries of the regular polytopes. 
In the following two sections we shall apply this framework to two
examples, namely the 16-cell, and then to the more complicated 24-cell.

\section{The ADHM 16-cell}\news \label{sec:16-cell}
As discussed earlier, the 16-cell has 8 vertices and hence the JNR bound
implies that the minimal charge for ADHM data with the symmetries of the
16-cell can be no greater than 7. 
In this section we enumerate and investigate all possibilities for
symmetric ADHM data with charge $N\le 7$, 
by a systematic consideration of all the possible
representations for  $Q_L$, $Q_R$, $q_R$ and $q_L,$
and the imposition of the ADHM constraint on the associated invariant data.

We have seen in Section \ref{sub16} that the symmetry group of the $16$-cell
 is generated by the left and right actions of the 
binary polyhedral group $\Q$, generated by $g_1 = i$ and $g_2 = j$ 
which satisfy 
\begin{equation}
g_1^2 = g_2^2 = (g_1 g_2)^2 = -1.
\end{equation}

\begin{figure}
\centering
\includegraphics[width=7cm]{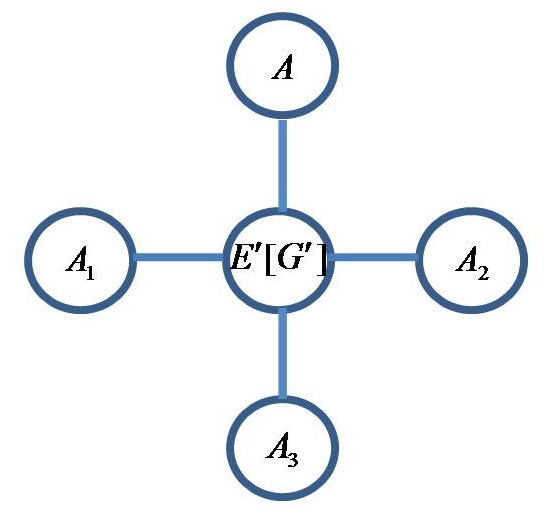}
\caption{
The Dynkin diagram of $\widetilde d_4$ providing a graphical
illustration of the irreducible representations of $\Q$ via the
McKay correspondence.}
\label{fig:d4}
\end{figure}

The representation theory of $\Q$ is captured by the Dynkin diagram
of $\widetilde d_4,$ presented in Figure~\ref{fig:d4}. 
This shows that there are four real 1-dimensional positive representations,
$A,A_1,A_2,A_3$ and a real 4-dimensional negative representation
$G',$ that is irreducible over $\R$ but reducible over $\C$ with the
decomposition $G'=E'\oplus E';$ recall our notation
$E'[G']$ to signify this. 

In a canonical basis we have
\begin{align} \label{eq:16-cell one dimensional reps}
A(g_1) &= 1, \quad \phantom{-_1} A(g_2) = 1, \\
A_1(g_1) &= -1, \quad A_1(g_2) = -1, \\
A_2(g_1) &= -1, \quad A_2(g_2) = 1, \\
A_3(g_1) &= 1, \quad \phantom{-} A_3(g_2) = -1,
\end{align}
and
\begin{equation} \label{eq:right basis of G}
G'(g_1) = \begin{pmatrix}
0 & -1 & 0 & 0 \\
1 & 0 & 0 & 0 \\
0 & 0 & 0 & -1 \\
0 & 0 & 1 & 0
\end{pmatrix}, \quad
G'(g_2) = \begin{pmatrix}
0 & 0 & -1 & 0 \\
0 & 0 & 0 & 1 \\
1 & 0 & 0 & 0 \\
0 & -1 & 0 & 0
\end{pmatrix}.
\end{equation}

There are four edges in the Dynkin diagram, corresponding to the 
four invariant maps between $G'\otimes E'$ and any of the 1-dimensional
representations. These four maps are $1\times 4$ matrices
that are easily obtained by solving (\ref{Brightaction}) and are
the building blocks of the ADHM data. Explicitly, the most
general invariant maps between $G'\otimes E'$ and $A,A_1,A_2,A_3$ are
given  by
\be
( 1, -i, -j, -k), \quad
( 1, i, j, -k), \quad
( 1, i, -j ,k ), \quad
( 1, -i, j, k),
\label{bblocks16}
\ee
respectively, 
where there is the freedom to multiply on the left by an arbitrary quaternion.

In terms of the notation introduced in our earlier framework,
the most general possibility is that
$Q^+_R=a_0 A \oplus a_1 A_1 \oplus a_2 A_2 \oplus a_3 A_3$, where
$a_0,a_1,a_2,a_3$ are non-negative integers (at least one
of which is non-zero) and
\begin{equation}
a_0 A = \underbrace{A \oplus \ldots \oplus A}_{\text{$a_0$ times}}.
\end{equation}
Furthermore, $Q_R^-=c_0G'$, for some positive integer $c_0.$

As we are only concerned with $N=\mbox{dim}(Q_R^+)+\mbox{dim}(Q_R^-)
=a_0+a_1+a_2+a_3+4c_0\le 7$
then immediately we see that $c_0=1,$ so that $Q_R^-=G'$ and
$a_0+a_1+a_2+a_3\le 3.$

Turning to the left action, $Q_L^+$ must also be the sum of some combination 
of $A$, $A_1$, $A_2$ and $A_3$, and $Q_L^-$ must be a copy of $G'$, 
but both blocks will generally be in a different basis to those of $Q_R$.
To find the left action, we must find representations in a basis which 
commute with the right action.
The matrices that commute with $Q_R^+$ are of the form 
$R_0 \oplus R_1 \oplus R_2 \oplus R_3$, where the $R_i$ are arbitrary 
square matrices of dimension $a_i$. 
We can perform an arbitrary basis transformation on each of these blocks 
without affecting the right action, and so can also write the left action 
in its irreducible form as the direct sum of some copies of 
$A$, $A_1$, $A_2$ and $A_3$, 
although not necessarily grouped together as in the right action.

The matrices that commute with $G'(g_1)$ and $G'(g_2)$ are of the form:
\begin{equation} \label{eq:form of matrices which commute with G in 16-cell}
\begin{pmatrix}
a & b & c & d \\
-b & a & -d & c \\
-c & d & a & -b \\
-d & -c & b & a
\end{pmatrix}.
\end{equation}
For a matrix in this form to square to $-\id_4$, 
it must satisfy $a = 0$, $b^2 + c^2 + d^2 = 1$. 
If we parameterise the two generators in the left action as
\begin{equation}
Q_L^-
(g_1) = 
\begin{pmatrix}
0 & b & c & d \\
-b & 0 & -d & c \\
-c & d & 0 & -b \\
-d & -c & b & 0
\end{pmatrix}, \quad 
Q_L^-
(g_2) = 
\begin{pmatrix}
0 & e & f & g \\
-e & 0 & -g & f \\
-f & g & 0 & -e \\
-g & -f & e & 0
\end{pmatrix},
\end{equation}
then the conditions for them to satisfy the group relations are
\begin{equation}
\begin{gathered}
b^2 + c^2 + d^2 = e^2 + f^2 + g^2 = 1, \\
be + cf + dg = 0.
\end{gathered}
\end{equation}
This is the condition that $(b,c,d)$ and $(e,f,g)$ are orthogonal 
unit vectors in $\R^3$. These can be rotated to be $(1, 0, 0)$ and 
$(0, 1, 0)$ by transformation matrices of the form 
\eqref{eq:form of matrices which commute with G in 16-cell}, which commute 
with the right action and so leave it invariant. The representation
$Q_L^-$ can therefore always be put in a basis where it has the following form
\begin{equation} \label{eq:left basis of G}
Q_L^-
(g_1) = 
\begin{pmatrix}
0 & 1 & 0 & 0 \\
-1 & 0 & 0 & 0 \\
0 & 0 & 0 & -1 \\
0 & 0 & 1 & 0
\end{pmatrix}, \quad 
Q_L^-(g_2) = 
\begin{pmatrix}
0 & 0 & 1 & 0 \\
0 & 0 & 0 & 1 \\
-1 & 0 & 0 & 0 \\
0 & -1 & 0 & 0
\end{pmatrix}.
\end{equation}
Note that this is the representation $Q_R^-=G'$ transformed by the matrix,
\begin{equation}
P = \diag(1, -1, -1, -1).
\end{equation}

Finally, $q_R$ and $q_L$ must each be one of the representations,
$2A,~2A_1,~2A_2,~\text{or}~2A_3,$
where there are always two copies of the same 1-dimensional representation 
in order to be a 1-dimensional quaternionic representation.

As we have $5\le N\le 7,$ 
this presents us with a finite, and reasonably small, number of 
possibilities. All have been investigated to find the most general 
left and right invariant maps, which are then tested to see 
if any also satisfy the ADHM constraint.
The result of this analysis is that there are no solutions with 
$N=5$ or $N=6,$ hence the
JNR bound is attained.

ADHM data is obtained for $N=7$ by taking $Q_R^+=Q_L^+=A\oplus A_2\oplus A_3$
and $q_R=q_L=2A_1.$
The symmetric data has the block structure
\begin{equation}
\M = \begin{pmatrix}
0 & 0 & 0 & L_1^{-} \\
0 & 0 & 0 & B_{11} \\
0 & 0 & 0 & B_{21} \\
0 & 0 & 0 & B_{31} \\
B_{11}^\T & B_{21}^\T & B_{31}^\T & 0
\end{pmatrix},
\label{data16}
\end{equation}
where 
\begin{align}
B_{11} &= b_1 ( 1, -i, -j, -k), \\
B_{21} &= b_2 ( 1, i, -j ,k ), \\
B_{31} &= b_3 ( 1, -i, j, k),\\
L_1^- &= l_0 ( 1, i, j, -k), 
\end{align}
with arbitrary real parameters $b_1,b_2,b_3,l_0.$
The right invariant building blocks (\ref{bblocks16})
are manifest in the invariant maps $B_{11},B_{21},B_{31}$ 
from $G'\otimes E'$ to $A,A_2,A_3,$ with left invariance reducing
the arbitrary quaternions to arbitrary real parameters.
As $q_R=2A_1$ then $L_1^-$ is a right invariant map from
from $G'\otimes E'$ to $2A_1$ and hence is formed from the
 remaining invariant building block in (\ref{bblocks16}),
where again left invariance reduces the
arbitrary quaternion to an arbitrary real parameter.

\begin{figure}
\centering
\includegraphics[width=5cm]{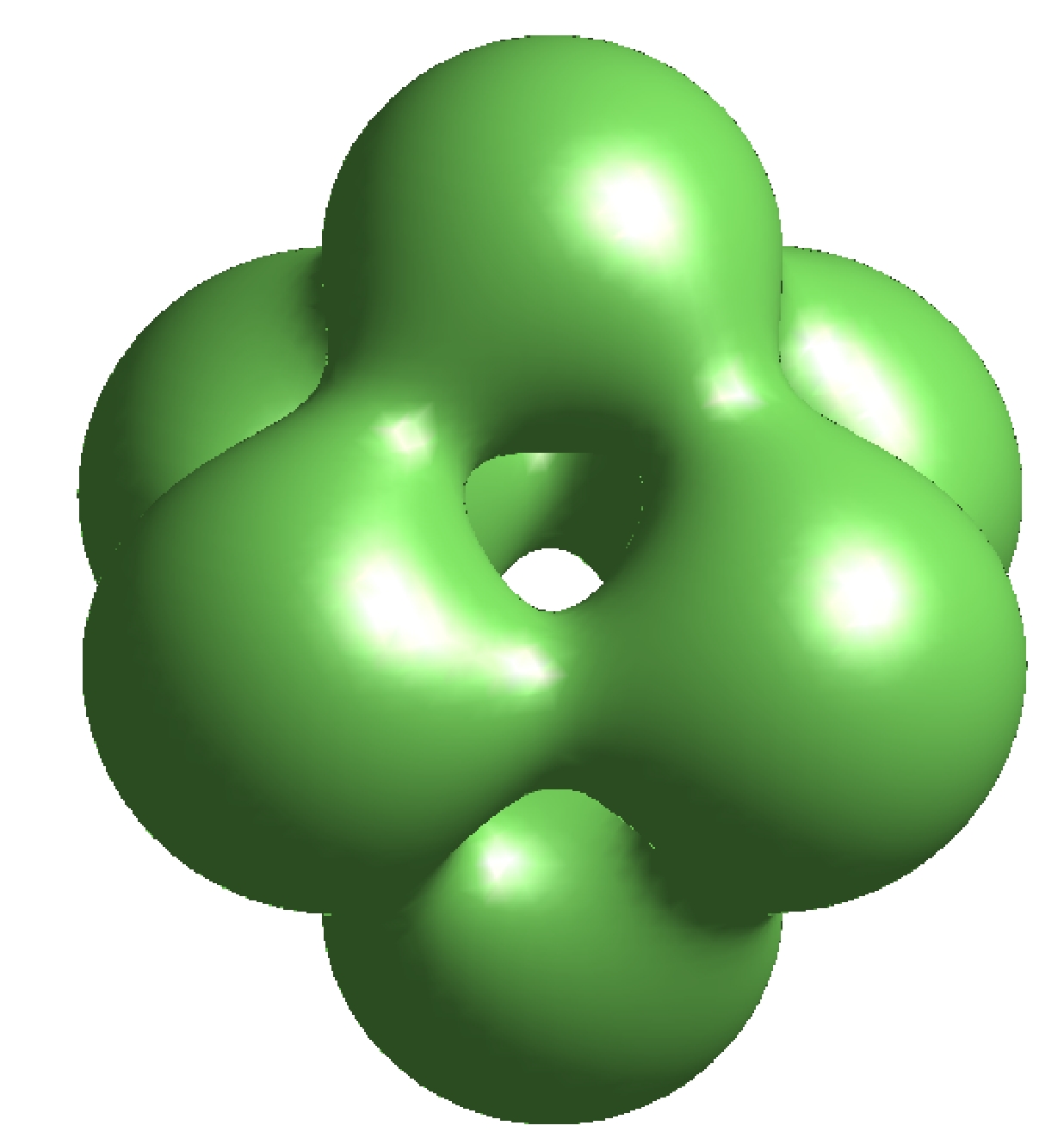}
\includegraphics[width=5cm]{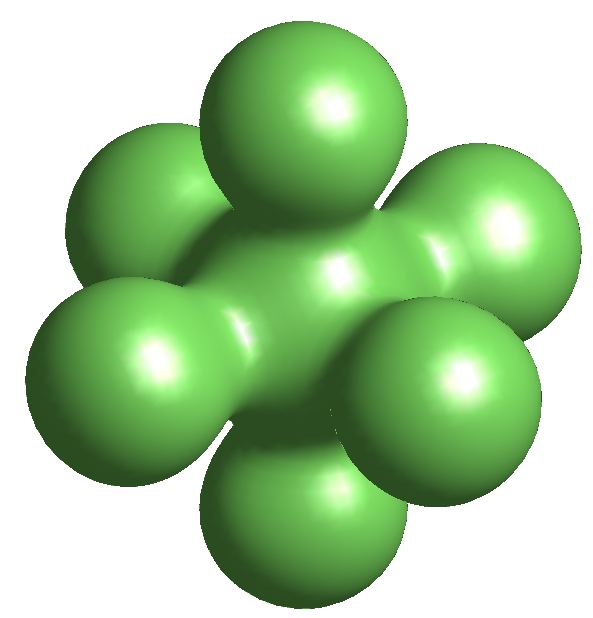}
\caption{
Surfaces of constant topological charge density for the
charge $7$ instanton with the symmetries of the $16$-cell. 
The left image is the charge density in the hyperplane
$x_4=0,$ where the vertices of the $16$-cell form an octahedron.
The right image is the charge density integrated along the $x_4$-direction. 
}
\label{fig:topological charge density of 16-cell instanton}
\end{figure}

The ADHM constraint applied to the data (\ref{data16}) reduces to the
equations
\begin{equation}
l_0^2 = b_1^2 = b_2^2 = b_3^3.
\end{equation}
Without loss of generality, we can take $l_0 = b_1 = b_2 = b_3 = \lambda$, 
with alternative choices of sign giving equivalent ADHM data. 
The remaining real parameter $\lambda$ is the arbitrary instanton scale.
Finally, we have the ADHM data of a charge 7 instanton with the symmetries 
of the 16-cell,
\begin{equation} \label{16cellADHM}
\M = \lambda \begin{pmatrix}
0 & 0 & 0 & 1 & i & j & -k \\
0 & 0 & 0 & 1 & -i & -j & -k \\
0 & 0 & 0 & 1 & i & -j & k \\
0 & 0 & 0 & 1 & -i & j & k \\
1 & 1 & 1 & 0 & 0 & 0 & 0 \\
-i & i & -i & 0 & 0 & 0 & 0 \\
-j & -j & j & 0 & 0 & 0 & 0 \\
-k & k & k & 0 & 0 & 0 & 0
\end{pmatrix}.
\end{equation}

The vertices of the 16-cell that lie in the 
$x_4 = 0$ hyperplane form an octahedron. 
The topological charge density (\ref{chargedensity}) of the 16-cell 
instanton in this hyperplane is plotted as an isosurface in the left image in 
Figure \ref{fig:topological charge density of 16-cell instanton}, 
using the formula (\ref{chargefromadhm}). The octahedron is clearly
visible in this image. The right image in  
Figure \ref{fig:topological charge density of 16-cell instanton}
captures more of the information about all the vertices by displaying
an isosurface of the charge density \edit{integrated} along the $x_4$-direction.

In the ADHM data presented above, the representation $A_1$ was distinguished
from the other three 1-dimensional representations.
However, any one of the 1-dimensional representations can be chosen as
the distinguished representation and this yields equivalent ADHM data.
In detail, all choices $Q_R^+= Q_L^+=A_i \oplus A_j \oplus A_k$ for 
$i,j,k\in\{0,1,2,3\}$, with $i \ne j \ne k$ are acceptable, where we have
used the notation $A_0 \equiv A$. With this choice then $q_R$ and 
$q_L$ may both be taken to be equal to two copies of the 1-dimensional
representation that is missing from $Q_R^+.$
 
As we have found a unique (up to scale) charge 7 instanton with the 
symmetries of the 16-cell then it must be equivalent to the JNR instanton
mentioned earlier. This is shown explicitly in Section \ref{sec:JNR}.

\section{The ADHM 24-cell}\news \label{sec:24-cell}
Our treatment of the 24-cell is similar to the 16-cell in the previous section,
upon replacing the binary dihedral group $\Q$ by the binary tetrahedral
group $\BT.$ 
 The main difference is that all but one of the real irreducible 
representations of the binary tetrahedral group have dimension greater than 
one, which makes finding appropriate commuting representations 
more complicated. 
Furthermore, the JNR bound in this case is $N\le 23,$ so we may need to
search up to charge 23.

We will first present the real irreducible representations of 
$\BT$
in some canonical basis, which can be taken as the basis 
for the representations in the right action, and this generates the 
ADHM building blocks associated with each edge in the Dynkin diagram. 
We then find the form that the representations in the left action must 
take in order to commute with the representations in the right action. 
Finally, we enumerate all the possible combinations of these 
representations up to charge 23 and test these to find ADHM data with the 
symmetries of the 24-cell.

\subsection{Representations of the right action of $\BT$} \label{sec:representations of the right action}

\begin{figure}
\centering
\includegraphics[width=11cm]{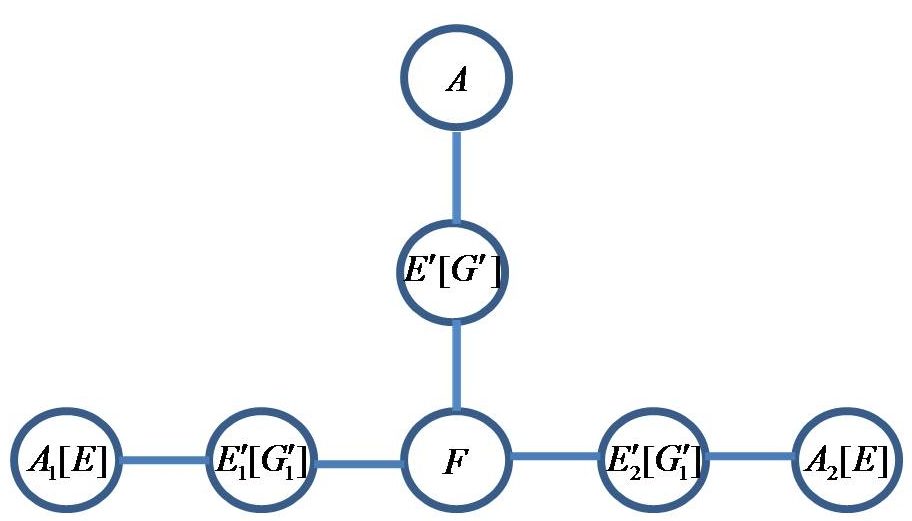}
\caption{
The Dynkin diagram of $\widetilde e_6$ providing a graphical
illustration of the irreducible representations of $\BT$ via the
McKay correspondence.}
\label{fig:e6}
\end{figure}

The representation theory of $\BT$ is captured by the Dynkin diagram of
 $\widetilde e_6,$ presented in Figure~\ref{fig:e6}.
There are real irreducible, over $\R,$ positive representations 
$A,E,F,$ but over $\C$ the 2-dimensional representation is 
reducible as $E=A_1\oplus A_2.$  In terms of the generators (\ref{genbt})
a canonical basis is 
\begin{gather}
A(g_1) = 1, \quad 
A(g_2) = 1, \\
E(g_1) = \frac{1}{2} \begin{pmatrix}
-1 & -\sqrt 3 \\
\sqrt 3 & -1
\end{pmatrix}, \quad
E(g_2) = \frac{1}{2} \begin{pmatrix}
-1 & \sqrt 3 \\
-\sqrt 3 & -1
\end{pmatrix}, \\
F(g_1) = \begin{pmatrix}
0 & 0 & 1 \\
1 & 0 & 0 \\
0 & 1 & 0 
\end{pmatrix}, \quad
F(g_2) = \begin{pmatrix}
0 & 1 & 0 \\
0 & 0 & -1 \\
-1 & 0 & 0
\end{pmatrix}.
\end{gather}
There are three complex negative representations, $E',E_1',E_2'$ which
\edit{combine} to form two real irreducible, over $\R,$ negative representations
$G'=E'\oplus E'$ and $G_1'=E_1'\oplus E_2'.$
A canonical basis for the generators is given by
\begin{equation}
G'(g_1) = \frac{1}{2} \begin{pmatrix}
1 & -1 & -1 & 1 \\
1 & 1 & -1 & -1 \\
1 & 1 & 1 & 1 \\
-1 & 1 & -1 & 1
\end{pmatrix}, \quad
G'(g_2) = \frac{1}{2} \begin{pmatrix}
1 & -1 & -1 & -1 \\
1 & 1 & 1 & -1 \\
1 & -1 & 1 & 1 \\
1 & 1 & -1 & 1
\end{pmatrix},
\end{equation}
\begin{equation}
\begin{gathered}
G'_1(g_1) = \frac{1}{4} \begin{pmatrix}
-1+\sqrt{3} & -1-\sqrt{3} & 1+\sqrt{3} & -1+\sqrt{3} \\
 1+\sqrt{3} & -1+\sqrt{3} & 1-\sqrt{3} & 1+\sqrt{3} \\
 -1+\sqrt{3} & -1-\sqrt{3} & -1-\sqrt{3} & 1-\sqrt{3} \\
 1+\sqrt{3} & -1+\sqrt{3} & -1+\sqrt{3} & -1-\sqrt{3}
\end{pmatrix},\\
G'_1(g_2) = \frac{1}{4} \begin{pmatrix}
 -1+\sqrt{3} & 1+\sqrt{3} & 1-\sqrt{3} & -1-\sqrt{3} \\
 -1-\sqrt{3} & -1+\sqrt{3} & 1+\sqrt{3} & 1-\sqrt{3} \\
 -1-\sqrt{3} & -1+\sqrt{3} & -1-\sqrt{3} & -1+\sqrt{3} \\
 1-\sqrt{3} & -1-\sqrt{3} & 1-\sqrt{3} & -1-\sqrt{3}
\end{pmatrix}.
\end{gathered}
\end{equation}

From the edges in the Dynkin diagram we see that there are four building
blocks associated with the invariant mappings 
from $G'\otimes E'$ to $A,\ $ 
from $G'\otimes E'$ to $F,\ $ 
from $G_1'\otimes E'$ to $F,\ $
and
from $G_1'\otimes E'$ to $E,\ $
where the last two blocks are both associated with two edges in the
Dynkin diagram since $G_1'$ contains both $E_1'$ and $E_2'$ in its
decomposition and $E$ contains both $A_1$ and $A_2$ in its
decomposition.

\subsection{Representations of the left action of $\BT$} \label{sec:representations of left action}

To find appropriate representations of the left action, we need to find a basis in which the representations commute with those given for the right action. 
Note that the right representation may include multiple copies of the same irreducible representation, such as $E \oplus E$. The corresponding left block may then be a $4 \times 4$ block rather than two separate $2 \times 2$ blocks, because the off-diagonal blocks may be non-zero here.

We will systematically go through the blocks in the representation of the right action and find the possible commuting representations of the left action. We will only consider the most granular blocks, so for example, if the right representation contains the $4 \times 4$ block $E \oplus E$, we would not consider the representation where the left block is also $E \oplus E$, since these both split into two $2 \times 2$ blocks. However, we will consider the representation where the left block is $E \otimes E$ since this is not composed of smaller blocks, and the whole $4 \times 4$ block must be considered together.

We recall that when we refer to a representation we mean the explicit 
matrices in Section \ref{sec:representations of the right action}. 
Likewise, when we use the tensor product and direct sum, we are 
referring to the concrete Kronecker product and direct sum of the 
matrices respectively.

Let us start with the trivial cases. If a block in the right action is simply the identity matrix then any positive representation of the appropriate size may be used as the block in left action. We can take these to be in the canonical basis since we can perform any basis transformation without affecting the form of the right block. Likewise, for any block in the right action which is a positive non-trivial representation, the block in the left action may be taken to be the identity matrix.

Consider the right representation $E$. The only matrices which commute with both $E(g_1)$ and $E(g_2)$ are of the form
\begin{equation}
\begin{pmatrix}
a & -b \\
b & a
\end{pmatrix}.
\end{equation}
These must be rotation matrices and the only non-trivial rotation 
matrices which form a representation of 
$\BT$ 
are $E(g_1)$ and $E(g_2)$. These two matrices are similar and the transformation between them is $P = \diag(1,-1)$, which does not commute with the right representation. So there are two possibilities for the left block: the original representation, $E$, and a twisted representation, $E^t$, where
\begin{equation}
E^t(g_1) = E(g_2), \quad E^t(g_2) = E(g_1).
\end{equation}

When the right representation is 
$2E \equiv E \oplus E \equiv \id_2 \otimes E$, the commuting matrices are of the form
\begin{equation} \label{eq:commuting with E+E}
\begin{pmatrix}
a & -b & c & -d \\
b & a & d & c \\
e & -f & g & -h \\
f & e & h & g
\end{pmatrix}.
\end{equation}
Here $E \otimes E$, $E \otimes E^t$ and $E \otimes 2A$ are possible representations for the left block. The twisted product, $E \otimes E^t$, is related to $E \otimes E$ via the transformation matrix $P = \diag(1,-1,1,-1)$. However, this does not leave the right action invariant and so $E \otimes E^t$ must be considered separately. Applying a twist to the first $E$ in the product can be undone since the transformation will apply only to the identity part of the right representation, $2E = \id_2 \otimes E$, and therefore leave it invariant. There is no need to consider $E \oplus E$ as a left representation, because both representations are then composed of smaller blocks that we have already considered.

It is not clear that these are all possible left representations for the right block $2E$. There may be other $4 \times 4$ matrices which are of the form in equation \eqref{eq:commuting with E+E} and form a representation but that are not related to $E \otimes E$ or $E \otimes E^t$ by a transformation which leaves the right action, $2E$, invariant. The condition for matrices of this form to be a representation is nonlinear and we have not been able to systemically rule out other possibilities. From now we will simply list possibilities for the left representations without claiming that these are exhaustive.

When the right representation is $3E$, 
the left representation must be $6$-dimensional and have an analogous
form to \eqref{eq:commuting with E+E} but generalised to a $6 \times 6$ matrix.
 Three such representations are $F \otimes E$, $F \otimes E^t$ and $F \otimes 2A$. We are free to choose the basis for $F$ since a transformation on the first term in the tensor product leaves the right block, $\id_3 \otimes E$, invariant. We will therefore take $F$ to be in the canonical basis above.

When the right representation is $4E$, let us start by considering the left representations in the form $\tilde G \otimes E$, where $\tilde G$ is some 
4-dimensional representation. These will commute with the right representation for any choice of $\tilde G$. We are free to choose a basis for $\tilde G$ without affecting the right representation, and so can always take it to be composed of irreducible blocks. There is no irreducible 4-dimensional positive representation, so in the appropriate basis $\tilde G \otimes E$ must be the 
direct sum of smaller blocks considered previously. Similarly, there is no need to consider left representations of the form $\tilde G \otimes E^t$ or $\tilde G \otimes 2A$.

We can also consider left representations in the form $\tilde E \otimes (E \oplus E)$, where $\tilde E$ is some 2-dimensional representation 
where we are free to choose the basis. The only choice for $\tilde E$ that does not decompose into smaller blocks is $\tilde E = E$, so that the left representation is $E \otimes (E \oplus E)$. By a similar argument, other possible left representations are of the form $E \otimes (\tilde E_1 \oplus \tilde E_2)$, where $\tilde E_1, \tilde E_2 = E, E^t$, or $2A$. Note that the ordering of the terms in the direct sum does not matter since these can be permuted without affecting the right representation.

There is no need to consider the left block in the form $\tilde E_1 \otimes \tilde E_2 \otimes \tilde E_3 \otimes \tilde E_4$ since this decomposes into blocks considered previously.

When the right representation is $5E$, there are no obvious possible 10-dimensional representations for the left representation which do not decompose into blocks we have already considered.

Following the same pattern, when the right representation is $6E$, the following left representations are possible and inequivalent: $F \otimes (\tilde E_1 \oplus \tilde E_2)$ and $E \otimes (\tilde E_1 \oplus \tilde E_2 \oplus \tilde E_3)$, where $E_1, E_2, E_3 = E, E^t$, or $2A$, and permutations of the direct sum are again equivalent. As before, if the left action is in the form $\tilde I \otimes \tilde E_1$ for some 6-dimensional representation $\tilde I,$ then it can be written as the sum of blocks considered previously, after the appropriate basis transformation.

This pattern also extends to the right representation being $7E$, $8E$ or $9E$, and the results are shown in Table \ref{table:representations}.

There are additional possibilities for the left representation when the right representation is $8E$. The matrices in $G'$ and $G_1'$ are all in the form 
 \eqref{eq:commuting with E+E} and so commute with $2E$. 
\edit{We can also consider the twisted representation, ${G'}^t,$ 
obtained by applying the transformation matrix 
\begin{equation} \label{eq:transformation between G' and G't}
P = \frac{1}{\sqrt 2} \begin{pmatrix}
1 & 0 & 0 & -1 \\
0 & 1 & 1 & 0 \\
0 & 1 & -1 & 0 \\
-1 & 0 & 0 & -1
\end{pmatrix}
\end{equation}
that swaps $G'(g_1)$ and $G'(g_2)$.
As this transformation commutes with $2E$ in the canonical basis, 
there is no need to consider ${G'}^t$ separately.
Similarly, there is a twisted representation ${G'_1}^t,$ 
however, the transformation between $G_1'$ and ${G_1'}^t$ does not commute with $2E$, so these must be considered separately. }
The representations $G' \otimes G'$, $G' \otimes G'_1$, $G' \otimes {G'_1}^t$, $G'_1 \otimes G'_1$, $G'_1 \otimes {G'_1}^t$ and $G'_1 \otimes G'$ are therefore also possible representations for the left representation when the right representation is $8E$. Note that these representations are positive as they are the tensor product of two negative representations.

There is no need to consider $10E$ or higher, since 
$Q_R^-$  must be at least 4-dimensional, and the highest charge that we 
need to consider is charge $23$.

The $F$ representation commutes only with the identity. If the right representation is $F$ there is therefore no non-trivial left representation.

When the right representation is $2F$, the only possible left representation is $E \otimes \id_3$.

When the right representation is $3F$, the only possible left representation is $F \otimes \id_3$.

For any higher dimensional right representation, $n F$, with $n > 3$, the left representation must be in the form $\rho_n \otimes \id_3$, where $\rho_n$ is 
an $n$-dimensional representation. However, we are free to choose the basis of $\rho_n$ and so can decompose it into irreducible representations, where each block has been considered previously.

The $G'_1$ representation commutes with matrices of the form 
\begin{equation}
\begin{pmatrix}
a & -b & 0 & 0 \\
b & a & 0 & 0 \\
0 & 0 & a & -b \\
0 & 0 & b & a
\end{pmatrix}.
\end{equation}
Neither $G'$ nor $G'_1$ can be put in this form since they are irreducible. For higher multiples of $G'_1$ in the right representation, the left representation must always occur with blocks of this form. For example, when the right representation is $2G'_1$, the left representation must be in the form
\begin{equation} \label{eq:tilde otimes}
\begin{pmatrix}
a & -b & 0 & 0 & c & -d & 0 & 0 \\
b & a & 0 & 0 & d & c & 0 & 0 \\
0 & 0 & a & -b & 0 & 0 & c & -d\\
0 & 0 & b & a & 0 & 0 & d & c\\
e & -f & 0 & 0 & g & -h & 0 & 0 \\
f & e & 0 & 0 & h & g & 0 & 0 \\
0 & 0 & e & -f & 0 & 0 & g & -h\\
0 & 0 & f & e & 0 & 0 & h & g
\end{pmatrix} \equiv
 \begin{pmatrix}
a & -b & c & -d \\
b & a & d & c \\
e & -f & g & -h \\
f & e & h & g
\end{pmatrix} \totimes \id_2,
\end{equation}
where we have defined $\totimes$ as the Kronecker product acting on each $2 \times 2$ block. We therefore see that $G' \totimes \id_2$, $G'_1 \totimes \id_2$ and ${G'_1}^t \totimes \id_2$ are possible left representations. The left representation ${G'}^t \totimes \id_2$ is equivalent to ${G'} \totimes \id_2$ since the transformation matrix between these is $P \totimes \id_2$, with $P$ as in 
 \eqref{eq:transformation between G' and G't} and so commutes with the right action.

There are no additional possibilities when the right representation is 
$3G'_1$ or $5G'_1$. There is no need to consider $6G'_1$ or higher as we would exceed charge $23$.

When the right representation is $4G'_1$, both $G' \otimes \tilde G$ and $G'_1 \otimes \tilde G$ are suitable left representations, where $\tilde G = \id_4$, $E \oplus E$ or $E^t \oplus E^t$.

The following left representations are also possible when the right block is $4G_1'$: $(E \otimes G') \totimes \id_2$, $(E \otimes {G'_1}) \totimes \id_2$, $(E \otimes {G'_1}^t) \totimes \id_2$, $(G' \otimes E) \totimes \id_2$, $(G' \otimes E^t) \totimes \id_2$, $(G'_1 \otimes E) \totimes \id_2$, and $(G'_1 \otimes E^t) \totimes \id_2$. 
The left representations in this form, where the first term is twisted,
 are related to the untwisted representations by transformations 
which do not affect the right action. Once again, the transformation between $(E \otimes G') \totimes \id_2$ and $(E \otimes {G'}^t) \totimes \id_2$ commutes with the right representation and so these do not need to be considered separately.

The final right representation to consider is $G'$, which commutes with matrices of the form
\begin{equation} \label{eq:commute with G'}
\begin{pmatrix}
a & -b & c & d \\
b & a & d & -c \\
-c & -d & a & -b \\
-d & c & b & a
\end{pmatrix}.
\end{equation}
The left representation can be $P G' P^\T$ or $P {G'}^t P^\T$, where $P = \diag(-1,1,1,1)$. We can see from the discussion above that $G'_1$ can never commute with $G'$ in any basis.

When the right representation is $2G'$, the left representation may be $E \otimes (P G' P^\T)$ or $E \otimes (P {G'}^t P^\T)$.

Similarly, when the right representation is $3G'$, the left representation may be $F \otimes P G' P^\T$ or $F \otimes P {G'}^t P^\T$.

When the right representation is $4G'$, both $G' \otimes \tilde G$ and $G'_1 \otimes \tilde G$ are suitable left representations, where $\tilde G$ is a positive representation in the form \eqref{eq:commute with G'}, $\tilde G  = 4A$, $E \oplus E$, or $E^t \oplus E^t$.

Any left representation of the form $\tilde G \otimes (P {G'}^t P^\T)$, 
can be decomposed into blocks which we have been 
considered previously by transforming $\tilde G$ to a basis where it is the direct sum of irreducible representations.

When the right representation is $5G'$, all possibilities are composed of blocks that we have previously considered. 

To recap, we have now
presented all possible blocks in the right representation which can 
appear up to charge $23$. For each block in the right representation,
 in the canonical basis, we have found possibilities for the 
corresponding
block in the left representation, many of which are actually the same representation but in a different basis. However, we cannot transform between these bases without affecting the right block and so we must consider these as inequivalent representations of the left action. 
A list of these possible representations is given in Table \ref{table:representations}. Unfortunately we have no method of systematically finding commuting representations and so we cannot rule out the possibility that there are other inequivalent left representations that we have not been able to find by inspection. 

{
\renewcommand{\arraystretch}{1.4}
\begin{table}
\centering
\begin{tabular}{p{0.27\textwidth} | p{0.67\textwidth}}
Right representation & Left Representation \\
\hline
$\id_n$	& $A$, $E$, or $F$ (for $n = 1,2,3$ respectively) \\
$E$		& $2A$, $E$, or $E^t$ \\
$2E$	& $E \otimes \tilde E_1$, where $\tilde E_1 = 2A$, $E$, or $E^t$. \\
$3E$	& $F \otimes \tilde E_1$ \\
$4E$	& $E \otimes (\tilde E_1 \oplus \tilde E_2)$ \\
$5E$	& --- \\
$6E$	& $F \otimes (\tilde E_1 \oplus \tilde E_2)$ or $E \otimes (\tilde E_1 \oplus \tilde E_2 \oplus \tilde E_3)$ \\
$7E$	& --- \\
$8E$	& $E \otimes (\tilde E_1 \oplus \tilde E_2 \oplus \tilde E_3 \oplus \tilde E_4)$, $G' \otimes \tilde G'$ or $G_1' \otimes \tilde G'$, where $\tilde G' = G', G_1'$ or ${G'_1}^t$.\\
$9E$	& $F \otimes (\tilde E_1 \oplus \tilde E_2 \oplus \tilde E_3)$ \\
$F$		& $\id_3$ \\
$2F$		& $E \otimes \id_3$ \\
$3F$		& $F \otimes \id_3$ \\
$n F$, $n > 3$	& --- \\
$G'$		& $P G' P^\T$, or $P {G'}^t P^\T$, where $P = \diag(-1,1,1,1)$. \\
$2G'$	& $E \otimes (P G' P^\T)$, or $E \otimes (P {G'}^t P^\T)$ \\
$3G'$	& $F \otimes (P G' P^\T)$, or $F \otimes (P {G'}^t P^\T)$ \\
$4G'$	& $G' \otimes \tilde G$, or $G_1' \otimes \tilde G$, where $\tilde G = 4A$, $E \oplus E$, or $E^t \oplus E^t$. \\
$5G'$	& --- \\
$G_1'$	& --- \\
$2G_1'$	& $G' \totimes \id_2$, $G'_1 \totimes \id_2$ or ${G'_1}^t \totimes \id_2$ \\
$3G_1'$	& --- \\
$4G_1'$	& $G' \otimes \tilde G$, or $G'_1 \otimes \tilde G$, where $\tilde G = \id_4$, $E \oplus E$ or $E^t \oplus E^t$; or $(E \otimes G') \totimes \id_2$, $(E \otimes {G'_1}) \totimes \id_2$, $(E \otimes {G'_1}^t) \totimes \id_2$, $(G' \otimes E) \totimes \id_2$, $(G' \otimes E^t) \totimes \id_2$, $(G'_1 \otimes E) \totimes \id_2$, or $(G'_1 \otimes E^t) \totimes \id_2$. \\
$5G_1'$	& ---
\end{tabular}
\caption{A summary of the possible blocks that make up the representations of the right and left actions of $\BT$ when acting on the ADHM data. }
\label{table:representations}
\end{table}
}

\subsection{A charge 23 solution}
Using computer algebra we have performed an \edit{automated and systematic
test of all tractable combinations} of the representations from 
the previous section to search for ADHM data up to charge 23.
This has resulted in a unique solution with charge 23, in which the right and 
left representations are
\begin{equation}
Q_R = E\oplus 3 F \oplus G' \oplus 2 G_1',
\end{equation}
and
\begin{equation}
Q_L =  E \oplus (F \otimes \id_3) \oplus (P G' P^\T) \oplus (G'_1 \totimes \id_2),
\end{equation}
where $P = \diag(-1,1,1,1)$ and $\totimes$ is the Kronecker product on $2 \times 2$ blocks as in \eqref{eq:tilde otimes}. In the block notation of our
framework, 
\begin{gather}
Q_{R,1}^+ = E, \; Q_{R,2}^+ = 3F, \; Q_{R,1}^- = G', \; Q_{R,2}^- = 2 G_1', \\
Q_{L,1}^+ = E, \; Q_{L,2}^+ = F \otimes \id_3, \; Q_{L,1}^- = P G' P^\T, \; Q_{L,2}^- = G_1' \totimes \id_2.
\end{gather}

The associated invariant blocks, which are again constructed from
the building blocks corresponding to the edges in the Dynkin diagram, are
\begin{equation}
\begin{split}
B_{12} &= b_1
\begin{pmatrix}
 -i & j & k & 1 & k & -1 & i & j \\
 -j & -i & -1 & k & 1 & k & -j & i
\end{pmatrix} \\
& \quad + b_2 \begin{pmatrix}
 j & i & 1 & -k & -1 & -k & j & -i \\
 -i & j & k & 1 & k & -1 & i & j
\end{pmatrix},
\end{split}
\end{equation}
and
\begin{equation}
B_{21} = b_3 \begin{pmatrix}
 1 & -i & j & -k \\
 -k & -j & -i & -1 \\
 j & -k & -1 & i \\
 k & -j & -i & 1 \\
 1 & i & -j & -k \\
 -i & 1 & -k & j \\
 -j & -k & 1 & i \\
 i & -1 & -k & j \\
 1 & i & j & k \\
\end{pmatrix},
\end{equation}
where $b_1, b_2$ and $b_3$ are arbitrary real coefficients, together with
$B_{22}$, which is presented in Figure \ref{fig:invariant matrix}.
Note that there is no invariant block $B_{11}$ as there is no
edge in the Dynkin diagram connecting $E$ to $G'.$
 
\begin{sidewaysfigure}
\centering
\renewcommand{\arraystretch}{1.5}
{ \footnotesize
\begin{equation*}
\begin{split}
B_{22} &= b_4 \begin{pmatrix}
 \frac{1}{2} \left(-i+\sqrt{3} j\right) & \frac{1}{2} \left(-\sqrt{3} i-j\right) & \frac{1}{2} \left(-\sqrt{3}-k\right) & \frac{1}{2} \left(1-\sqrt{3} k\right) & \frac{1}{2} \left(\sqrt{3}-k\right) & \frac{1}{2} \left(-1-\sqrt{3} k\right) & \frac{1}{2} \left(i+\sqrt{3} j\right) & \frac{1}{2} \left(\sqrt{3} i-j\right) \\
 j & -i & 1 & k & 1 & -k & -j & -i \\
 \frac{1}{2} \left(\sqrt{3}-k\right) & \frac{1}{2} \left(1+\sqrt{3} k\right) & \frac{1}{2} \left(i+\sqrt{3} j\right) & \frac{1}{2} \left(-\sqrt{3} i+j\right) & \frac{1}{2} \left(i-\sqrt{3} j\right) & \frac{1}{2} \left(-\sqrt{3} i-j\right) & \frac{1}{2} \left(\sqrt{3}+k\right) & \frac{1}{2} \left(1-\sqrt{3} k\right) \\
 j & -i & -1 & -k & -1 & k & -j & -i \\
 \frac{1}{2} \left(i+\sqrt{3} j\right) & \frac{1}{2} \left(-\sqrt{3} i+j\right) & \frac{1}{2} \left(\sqrt{3}-k\right) & \frac{1}{2} \left(1+\sqrt{3} k\right) & \frac{1}{2} \left(-\sqrt{3}-k\right) & \frac{1}{2} \left(-1+\sqrt{3} k\right) & \frac{1}{2} \left(-i+\sqrt{3} j\right) & \frac{1}{2} \left(\sqrt{3} i+j\right) \\
 \frac{1}{2} \left(1-\sqrt{3} k\right) & \frac{1}{2} \left(\sqrt{3}+k\right) & \frac{1}{2} \left(\sqrt{3} i+j\right) & \frac{1}{2} \left(-i+\sqrt{3} j\right) & \frac{1}{2} \left(-\sqrt{3} i+j\right) & \frac{1}{2} \left(i+\sqrt{3} j\right) & \frac{1}{2} \left(-1-\sqrt{3} k\right) & \frac{1}{2} \left(-\sqrt{3}+k\right) \\
 \frac{1}{2} \left(-\sqrt{3}-k\right) & \frac{1}{2} \left(-1+\sqrt{3} k\right) & \frac{1}{2} \left(i-\sqrt{3} j\right) & \frac{1}{2} \left(-\sqrt{3} i-j\right) & \frac{1}{2} \left(i+\sqrt{3} j\right) & \frac{1}{2} \left(-\sqrt{3} i+j\right) & \frac{1}{2} \left(-\sqrt{3}+k\right) & \frac{1}{2} \left(-1-\sqrt{3} k\right) \\
 \frac{1}{2} \left(-1-\sqrt{3} k\right) & \frac{1}{2} \left(-\sqrt{3}+k\right) & \frac{1}{2} \left(-\sqrt{3} i+j\right) & \frac{1}{2} \left(i+\sqrt{3} j\right) & \frac{1}{2} \left(\sqrt{3} i+j\right) & \frac{1}{2} \left(-i+\sqrt{3} j\right) & \frac{1}{2} \left(1-\sqrt{3} k\right) & \frac{1}{2} \left(\sqrt{3}+k\right) \\
 -i & j & -k & -1 & -k & 1 & i & j \\
\end{pmatrix} \\
& \qquad + b_5 \begin{pmatrix}
 \frac{1}{2} \left(\sqrt{3} i+j\right) & \frac{1}{2} \left(-i+\sqrt{3} j\right) & \frac{1}{2} \left(-1+\sqrt{3} k\right) & \frac{1}{2} \left(-\sqrt{3}-k\right) & \frac{1}{2} \left(1+\sqrt{3} k\right) & \frac{1}{2} \left(\sqrt{3}-k\right) & \frac{1}{2} \left(-\sqrt{3} i+j\right) & \frac{1}{2} \left(i+\sqrt{3} j\right) \\
 i & j & -k & 1 & k & 1 & i & -j \\
 \frac{1}{2} \left(-1-\sqrt{3} k\right) & \frac{1}{2} \left(\sqrt{3}-k\right) & \frac{1}{2} \left(\sqrt{3} i-j\right) & \frac{1}{2} \left(i+\sqrt{3} j\right) & \frac{1}{2} \left(\sqrt{3} i+j\right) & \frac{1}{2} \left(i-\sqrt{3} j\right) & \frac{1}{2} \left(-1+\sqrt{3} k\right) & \frac{1}{2} \left(\sqrt{3}+k\right) \\
 i & j & k & -1 & -k & -1 & i & -j \\
 \frac{1}{2} \left(\sqrt{3} i-j\right) & \frac{1}{2} \left(i+\sqrt{3} j\right) & \frac{1}{2} \left(-1-\sqrt{3} k\right) & \frac{1}{2} \left(\sqrt{3}-k\right) & \frac{1}{2} \left(1-\sqrt{3} k\right) & \frac{1}{2} \left(-\sqrt{3}-k\right) & \frac{1}{2} \left(-\sqrt{3} i-j\right) & \frac{1}{2} \left(-i+\sqrt{3} j\right) \\
 \frac{1}{2} \left(-\sqrt{3}-k\right) & \frac{1}{2} \left(1-\sqrt{3} k\right) & \frac{1}{2} \left(i-\sqrt{3} j\right) & \frac{1}{2} \left(\sqrt{3} i+j\right) & \frac{1}{2} \left(-i-\sqrt{3} j\right) & \frac{1}{2} \left(-\sqrt{3} i+j\right) & \frac{1}{2} \left(\sqrt{3}-k\right) & \frac{1}{2} \left(-1-\sqrt{3} k\right) \\
 \frac{1}{2} \left(1-\sqrt{3} k\right) & \frac{1}{2} \left(-\sqrt{3}-k\right) & \frac{1}{2} \left(\sqrt{3} i+j\right) & \frac{1}{2} \left(i-\sqrt{3} j\right) & \frac{1}{2} \left(\sqrt{3} i-j\right) & \frac{1}{2} \left(i+\sqrt{3} j\right) & \frac{1}{2} \left(1+\sqrt{3} k\right) & \frac{1}{2} \left(-\sqrt{3}+k\right) \\
 \frac{1}{2} \left(\sqrt{3}-k\right) & \frac{1}{2} \left(-1-\sqrt{3} k\right) & \frac{1}{2} \left(-i-\sqrt{3} j\right) & \frac{1}{2} \left(-\sqrt{3} i+j\right) & \frac{1}{2} \left(i-\sqrt{3} j\right) & \frac{1}{2} \left(\sqrt{3} i+j\right) & \frac{1}{2} \left(-\sqrt{3}-k\right) & \frac{1}{2} \left(1-\sqrt{3} k\right) \\
 -j & -i & 1 & -k & -1 & -k & -j & i \\
\end{pmatrix}.
\end{split}
\end{equation*}
}
\caption{The invariant map from $G'_1\otimes E'$ to $3F$.
The coefficients $b_4$ and $b_5$ are real and arbitrary.}
\label{fig:invariant matrix}
\end{sidewaysfigure}
\begin{figure}
\centering
\includegraphics[width=10cm]{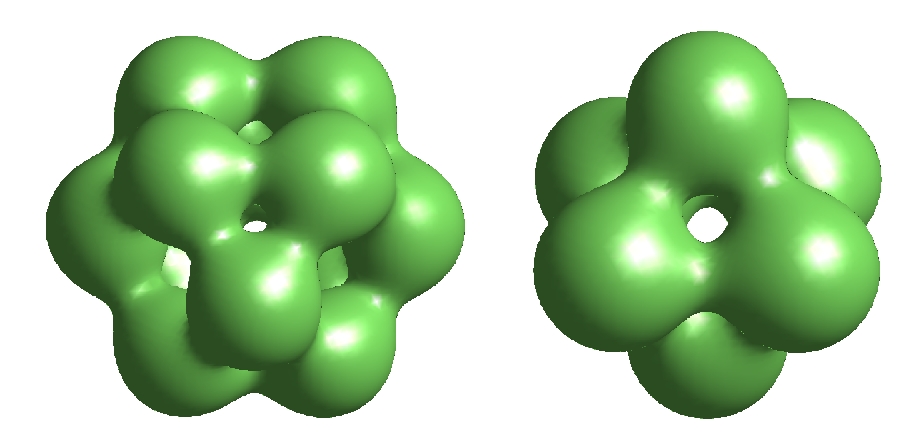}
\includegraphics[width=5cm]{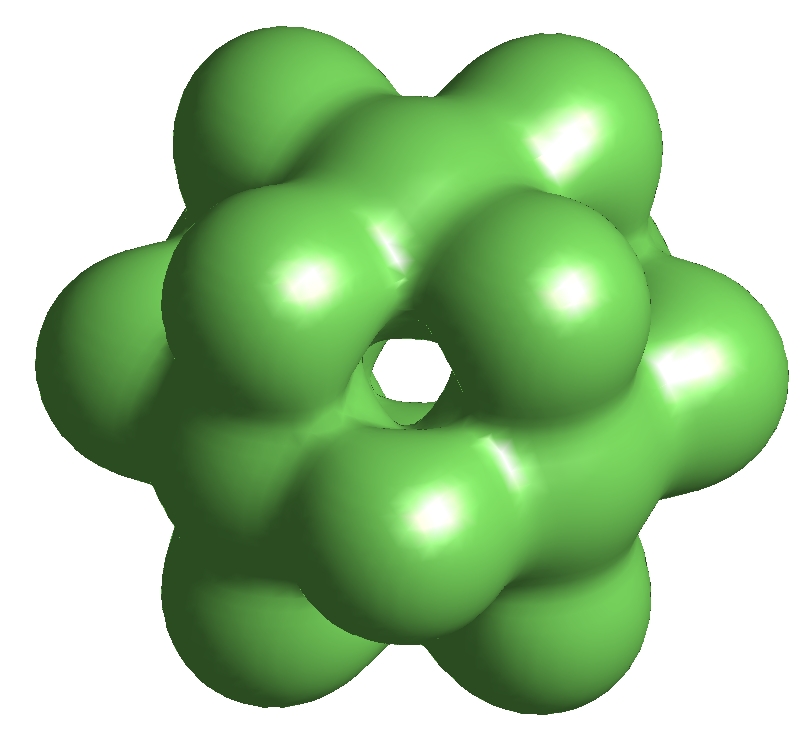}
\caption{
Surfaces of constant topological charge density for the
charge $23$ instanton with the symmetries of the $24$-cell. 
The left image is the charge density in the hyperplane
$x_4=0,$ where the vertices of the $24$-cell form a cuboctahedron.
The centre image is the charge density in the hyperplane
$x_4=1,$ where the vertices of the $24$-cell form an octahedron.
The right image is the charge density integrated along the $x_4$-direction.
}
\label{fig:24 cell cuboctahedron}
\end{figure}

The quaternionic representations are $q_R=q_L=2A.$
As there is no edge in the Dynkin diagram connecting $G_1'$ to 
$A$ then $L_2^-=0$ and the only non-vanishing block in $L$ is 
the invariant $L_1^-,$ which exists because there is an edge in
the Dynkin diagram connecting $G'$ to $A.$
Explicitly, this right and left invariant row vector is
\begin{equation}
L_{1}^- = l_1 (1, -i , -j , -k),
\end{equation}
where $l_1$ is an arbitrary real parameter.

The ADHM data assembled from these invariant blocks is then
\begin{equation} \label{24cellADHM}
\M = \begin{pmatrix}
0 & 0 & L_1^- & 0 \\
0 & 0 & 0 & B_{12} \\
0 & 0 & B_{21} & B_{22} \\
0 & B_{21}^\T & 0 & 0 \\
B_{12}^\T & B_{22'}^\T & 0 & 0
\end{pmatrix}.
\end{equation}
Applying the 
ADHM condition yields the following constraints on the coefficients
\begin{equation}
3(b_4^2 + b_5^2) = 3 b_3^2 = 2(b_1^2 + b_2^2) = l_1^2.
\end{equation}
These can be solved with the parameterisation
\begin{equation}
\begin{gathered}
b_1 = \frac{\lambda}{2\sqrt{2}} \cos \theta_1, \quad
b_2 = \frac{\lambda}{2\sqrt{2}} \sin \theta_1, \quad
b_3 = \frac{\lambda}{2\sqrt{3}}, \\
b_4 = \frac{\lambda}{2\sqrt{3}} \cos \theta_2, \quad
b_5 = \frac{\lambda}{2\sqrt{3}} \sin \theta_2, \quad
l_1 = \frac{\lambda}{2},
\end{gathered}
\end{equation}
where any choice of the parameters $\theta_1$ and $\theta_2$ gives 
equivalent ADHM data.
The overall scale is given by $\lambda$ and
\begin{equation}
\M^\dagger \M = \lambda^2\,  \id_{23}.
\end{equation}

The vertices of the 24-cell can be divided into three hyperplanes, so
that in the first of these hyperplanes the vertices form a cuboctahedron 
(a cube with each corner cut off to give an equilateral triangle face).
The vertices in the two remaining hyperplanes form octahedrons.
Figure \ref{fig:24 cell cuboctahedron} displays surfaces of constant 
topological charge density obtained from the above ADHM data.
The first image is in the $x_4 = 0$ hyperplane, where the cuboctahedral 
structure is clear. The second image is in the $x_4 = 1$ hyperplane where the 
octahedral structure is clear. Finally, the third image is obtained by
integrating the topological charge density along the $x_4$-direction
and reveals a merged version of the two structures.

The ADHM data that we have found with the 
symmetries of the 24-cell has a charge equal to that
given by the JNR bound and is therefore expected to be 
equivalent to a JNR instanton. 
We shall address this issue in the following section.

\section{Equivalence to JNR data}\news \label{sec:JNR}
The three examples of ADHM data that we have computed for the
5-cell, 16-cell and 24-cell all have a charge equal to the JNR bound.
The ADHM data should therefore be equivalent to JNR data in which 
points with equal weights are placed at the vertices of these polytopes.
In this section we shall explicitly demonstrate this equivalence.  

The ADHM data corresponding to general JNR data 
has been presented in \cite{CFTG}, but in a different format to
the canonical form of ADHM data given by (\ref{standardform}).
For charge $N$ JNR data with equal weights and points in $\R^4$ given
by $y_0, \ldots, y_N,$ the ADHM data is \cite{CFTG}
\begin{align}
\Delta(x) &= \begin{pmatrix}
y_0 & \cdots & y_0 \\
y_1 & &  \\
& \ddots &  \\
& & y_N
\end{pmatrix} - 
\begin{pmatrix}
1 & \cdots & 1 \\
1 & & \\
& \ddots & \\
& & 1
\end{pmatrix} x. 
\end{align}
To convert this ADHM data to the canonical form (\ref{standardform}) 
we need matrices $S \in \O(N+1)$ and $C \in \GL(N, \R)$ such that
\begin{equation}
\begin{pmatrix}
0 & \cdots & 0 \\
1 & & \\
& \ddots & \\
& & 1
\end{pmatrix} =
S \begin{pmatrix}
1 & \cdots & 1 \\
1 & & \\
& \ddots & \\
& & 1
\end{pmatrix} C.
\end{equation}
For general charge $N,$ the following matrices will perform this 
transformation
\begin{equation}
C_{ij} = \begin{cases}
0 & \text{if $i > j$} \\
{\displaystyle \frac{j}{\sqrt{j(j+1)}}} & \text{if $i = j$} \\
{\displaystyle -\frac{1}{\sqrt{j(j+1)}}} & \text{if $ i < j$}
\end{cases} \qquad \text{where} \quad i,j = 1, \ldots, N,
\end{equation}
and
\begin{equation}
S = \begin{pmatrix}
- \frac{1}{\sqrt{N+1}} & \begin{matrix} \frac{1}{\sqrt{N+1}} & \cdots & \frac{1}{\sqrt{N+1}} \end{matrix} \\
\begin{matrix}
\ \ C_{1 1}  \\ 
-C_{1 2}  \\
\vdots \\
-C_{1 N}
\end{matrix} & \del{C^\T}_{ij}
\end{pmatrix}.
\end{equation}
This is a generalisation of the transformation presented in 
\cite{Os} for $N = 1,2$.

In the case of the 5-cell with $N=4,$  
the points are taken to be the five vertices
\begin{equation}
\begin{gathered}
y_0 = \frac{1}{4} \del { 1 - \sqrt 5 \del{i + j + k} }, \quad
y_1 = \frac{1}{4} \del { 1 - \sqrt 5 \del{i - j - k} }, \\
y_2 = \frac{1}{4} \del { 1 - \sqrt 5 \del{-i + j - k} }, \quad
y_3 = \frac{1}{4} \del { 1 - \sqrt 5 \del{-i - j + k} }, 
\quad\quad
y_4 = -1.
\end{gathered}
\end{equation}
The ADHM 5-cell data $\M$ presented earlier in (\ref{5cellADHM}) 
is equivalent to this data, when $\lambda=\frac{1}{4},$ since 
\begin{equation}
\M = \begin{pmatrix}
1 & 0 \\
0 & Q
\end{pmatrix}
S
\begin{pmatrix}
y_0 & \cdots & y_0 \\
y_1 & &  \\
& \ddots &  \\
& & y_4
\end{pmatrix}
C \,
Q^{-1},
\end{equation}
where $S$ and $C$ are given above, and
\begin{equation}
Q = \begin{pmatrix}
 0 & 0 & 0 & 1 \\
 0 & -\frac{\sqrt 2}{\sqrt 3} & -\frac{1}{\sqrt{3}} & 0 \\
 -\frac{1}{\sqrt{2}} & \frac{1}{\sqrt{6}} & -\frac{1}{\sqrt{3}} & 0 \\
 -\frac{1}{\sqrt{2}} & -\frac{1}{\sqrt{6}} & \frac{1}{\sqrt{3}} & 0 \\
\end{pmatrix}.
\end{equation}
Of course, the JNR data can be scaled to provide 
equivalence for any value of the scale $\lambda.$

To construct charge 7 JNR data with the symmetries of the 16-cell, 
we can take the 8 points to be the vertices of the 16-cell 
\begin{equation}
\begin{gathered}
y_0 = 1,\; y_1 = -1,\; y_2 = i,\; y_3 = -i, \\
y_4 = j,\; y_5 = -j,\; y_6 = k,\; y_7 = -k.
\end{gathered}
\end{equation}
The ADHM data constructed previously,
(\ref{16cellADHM}) with $\lambda=\frac{1}{2}$,
is equivalent to this JNR data using the same transformations as
above with
\begin{equation}
Q = \begin{pmatrix}
 0 & -\frac{1}{\sqrt{3}} & -\frac{1}{\sqrt{6}} & -\frac{1}{\sqrt{10}} & -\frac{1}{\sqrt{15}} & \frac{2}{\sqrt{21}} & \frac{1}{\sqrt{7}} \\
 0 & 0 & 0 & -\frac{\sqrt 2}{ \sqrt 5} & -\frac{2}{\sqrt{15}} & -\frac{2}{\sqrt{21}} & -\frac{1}{\sqrt{7}} \\
 0 & -\frac{1}{\sqrt{3}} & -\frac{1}{\sqrt{6}} & \frac{1}{\sqrt{10}} & \frac{1}{\sqrt{15}} & -\frac{2}{\sqrt{21}} & -\frac{1}{\sqrt{7}} \\
 -1 & 0 & 0 & 0 & 0 & 0 & 0 \\
 0 & \frac{1}{\sqrt{3}} & -\frac{\sqrt 2}{\sqrt 3} & 0 & 0 & 0 & 0 \\
 0 & 0 & 0 & \frac{\sqrt 2}{\sqrt 5} & -\frac{\sqrt 3}{ \sqrt 5} & 0 & 0 \\
 0 & 0 & 0 & 0 & 0 & -\frac{\sqrt 3}{ \sqrt 7} & \frac{2}{\sqrt{7}}
\end{pmatrix}.
\end{equation}

Due to the large dimension of the charge 23 ADHM data, 
it is difficult to find the transformation matrix between the 
solution in (\ref{24cellADHM}) and the JNR generated ADHM data. 
However, by examining the eigenvalues of the matrices $Q_L$ and $Q_R$ which 
leave the JNR generated ADHM data invariant under the left and right action 
of $\BT$, we are able to confirm that they are the same representations as 
appear in the earlier ADHM data (\ref{24cellADHM}). 

\section{Discussion and conclusion}\news
In this paper we have understood how the ADHM data of a 
charge $N$ symmetric instanton transforms under the action of 
a finite subgroup of $SO(4).$
Given the description of the ADHM data in terms of quaternions, 
the natural way to represent the action of such a symmetry group is via 
the lift to the double cover, which is a subgroup of 
$SU(2) \times SU(2)$ and acts via right and left multiplication
by unit quaternions. 
For the symmetry group of the 5-cell, the double cover is isomorphic to a 
subgroup of $SU(2)$, and the left and right actions are not independent. 
For this action, with elements $(g^\sharp, g)$, 
where $g^\sharp$ is dependent on $g$, the ADHM data transforms under a single 
$N$-dimensional real representation, $Q$, and a single 1-dimensional 
quaternionic representation, $q$. 
These may always be taken to be in the canonical basis where $Q$ is the 
direct sum of irreducible representations. It is then straightforward 
to enumerate all combinations of irreducible representations and search 
for any ADHM data that is invariant.
This procedure allowed us to construct the ADHM data of a charge 4 
instanton with the symmetries of the 5-cell, and show that this is 
lowest charge instanton with these symmetries.

The double cover of the symmetry groups of the 
remaining polytopes take the form
$\BG\times\BG,$ where $\BG$ is one of the binary polyhedral groups
$\Q, \BT$ or $\BY,$ 
and the left and right actions of these groups are independent. 
This means that there are two independent representations of $\BG$, 
$Q_R$ and $Q_L,$ and we only have the freedom to choose a basis in which 
either $Q_R$ or $Q_L$ is explicitly the direct sum of irreducible 
representations. However, $Q_R$ and $Q_L$ must commute, 
so the possible form of the representation $Q_L$ is restricted when 
$Q_R$ is in the canonical basis. In the case of the 16-cell, this has 
allowed us to uniquely determine all possibilities for $Q_L$ given a 
choice of $Q_R$. For the 24-cell, we have only been able to determine the 
nonlinear constraints on the form of the representations in $Q_L$, 
and find the obvious examples by inspection.

With all possible combinations of $Q_R$ and $Q_L$ known for the 16-cell, 
and a large number known for the 24-cell, we have tested each combination 
to determine if there is invariant data that 
also satisfies the ADHM constraint. 
For the 16-cell we have found a solution at charge 7 and for
the 24-cell we have found a solution at charge 23, 
both of which have been shown to be equivalent to JNR data.

In previous work on instantons with platonic symmetries, the minimal
charge instantons associated with the cube and dodecahedron are not of
the JNR type, and perhaps this is related to the fact that they are not 
deltahedra. 
In our search for instantons with polytope symmetries, the three minimal
charge examples we have constructed are all of the JNR type, and perhaps
the explanation again lies in the fact that the 5-cell, 16-cell and 24-cell
all have triangular faces. This suggests that the 8-cell and the 120-cell
may be more promising candidates to find minimal charge instantons that
are not of the JNR type. However, the JNR bound for the 120-cell is
$N\le 599,$ which is clearly beyond the limits of our approach. 
For the 8-cell, the JNR bound is $N\le 15$ and this is also at the limits 
of our capabilities because there are four 1-dimensional representations
and this rapidly generates a large number of possibilities as the charge
increases, and in particular produces invariant data with too many parameters
to make the ADHM constraint tractable.
 
As the 8-cell is dual to the 16-cell then they share the same symmetry
group, so it might be tempting to conclude from our analysis that there
is no instanton associated with the 8-cell with charge less than 8.
However, there are a number of caveats to this conclusion, as we now discuss.

In the case of platonic symmetry, a polyhedral group acts 
as spatial rotations and there is an action on the gauge potential that
covers this, but potentially the image of this representation may only
be a quotient of the polyhedral group, rather than the full 
polyhedral group itself.
If this is the case, then in passing to the binary polyhedral group, 
as is natural for the quaternionic ADHM description, there will be a double 
cover of this quotient group, but this may not be equal to some quotient of
the binary polyhedral group \cite{SiSu}. Precisely this situation occurs
for the minimal charge instanton associated with the cube, and as the
8-cell is the 4-dimensional analogue of the cube then perhaps 
something similar might occur, taking the 8-cell outside our framework.
 
It is also possible that lower charge solutions exist, but outside of our 
framework, for the following reasons. 
In the 24-cell, there may be representations in the left action that we have 
not identified and yield a lower charge solution. 
To rule out this possibility would require the general solution of a set
of nonlinear constraints to find the most general form of representations 
that commute with any given right representation, 
and it is not clear how to proceed with this.
Our framework was therefore restricted to identifying
obvious low-dimensional commuting representations and using these to 
form larger representations by forming tensor products.
Some evidence to support the validity of this approach is the fact that
we were able to obtain the charge 23 solution through this mechanism, which
has a fairly complicated structure for both the left and right representations.

We have also assumed that both $Q_R$ and $Q_L$ form representations of the 
appropriate binary polyhedral group. 
It is possible that there are symmetric instantons with ADHM data that
 is invariant under some matrices $Q_R$ and $Q_L$ which are not strictly 
representations. 
For example, consider the right action of $g_i^2 = -1$ in the double 
cover of the 16-cell symmetry group. Then there must exist matrices, 
$Q_R(g_i)$, such that
\begin{equation}
(Q_R(g_i))^2 M = - M (Q_R(g_i))^2.
\end{equation}
If $Q_R$ is composed of irreducible representations then we saw previously 
that in the appropriate basis
$(Q_R(g_i))^2 = \diag(\id_m, -\id_n).$ 
 However, if $N$ is even then the following is also a possibility,
\begin{equation}
(Q_R(g_i))^2 = \begin{pmatrix}
0 & \id_{N/2} \\
-\id_{N/2} & 0
\end{pmatrix},
\end{equation}
where $M$ takes the form
\begin{equation}
M = \begin{pmatrix}
A & B \\
B & A
\end{pmatrix},
\end{equation}
with $A$ and $B$ symmetric matrices. 
The matrices $Q_R(g_i)$ do not form a representation
of $\Q,$ for example $g_1^4 = 1$, yet $(Q_R(g_1))^4 = - \id_N$.
However, $Q_R(g_i)$
still obey the group action when applied to 
$M$ since the sign is projected out.
We have not been able to construct an argument why this cannot occur, though
one may indeed exist.

Another possibility is that $Q_R$ and $Q_L$ are representations of opposite sign. We took $Q_R$ to be composed of positive representations in the upper block and negative representations in the lower block, so that 
$(Q_R(g_i))^\alpha = \diag(\id_m, -\id_n)$. 
We also took a similar block structure for $Q_L$, but it is possible that $Q_L$ consists of negative representations in the upper block and positive representations in the lower block so that 
$(Q_L(g_i))^\alpha = \diag(- \id_m, \id_n)$. 
Again, this difference of sign is irrelevant in the action on the ADHM data. 
We have performed a similar analysis as in Section \ref{sec:representations of left action} with the left representations having the opposite sign, but we were not able to find any invariant data of this form. Again, as for the representations of the same sign, our search was not exhaustive, but there may be some
simple argument that rules out this possible structure.

Finally, it is possible that the matrices $Q_R$ and $Q_L$ only satisfy the group presentation up to a sign,
\begin{equation}
(Q_{R,L}(g_1))^\alpha = \pm (Q_{R,L}(g_2))^\beta = \pm \del{ Q_{R,L}(g_1) Q_{R,L}(g_2) }^\gamma.
\end{equation}
For the 24-cell symmetry group, where $\alpha = \beta  = 3$, we can always choose the sign of $Q_{R,L}(g_i)$ such that the signs in this expression match. However, for the 16-cell, where $\alpha = \beta = 2$, it 
\edit{may} be possible to have symmetric ADHM data which is invariant under some matrices $Q_{R,L}$ where the signs do not match. This would not be equivalent to a true representation.
The core problem that generates all these possibilities outside of our
framework 
is that the transformation of the ADHM data is unaffected by the sign of 
$Q_{R,L}$  and so they need only satisfy the group operation up to a sign,
\begin{equation}
Q_{R,L}(g) Q_{R,L}(h) = \pm Q_{R,L}(gh).
\end{equation}
Our treatment in terms of representation theory
 is therefore only applicable when 
the signs agree with the group operation. 
We have been unable to find meaningful examples of suitable matrices when 
the signs do not agree.

\section*{Acknowledgements}
JPA is supported by an STFC studentship.
PMS acknowledges funding from EPSRC under grant EP/K003453/1 and 
STFC under grant ST/J000426/1.

\end{document}